\input harvmac
\noblackbox
\ifx\answ\bigans
\magnification=1200\baselineskip=14pt plus 2pt minus 1pt
\else\baselineskip=16pt 
\fi

\def\Cb{\ov C}
\def\cx#1{{\cal{#1}}}

\def\yb{\ov y}\def\Ga{\Gamma}\def\bb#1{\ov #1}
\def\rr{\rangle}\def\ll{\langle}\def\tx#1{\tilde{#1}}
\def\tb{type $IIB$\ }\def\ta{type $IIA$\ }\def\ti{type $I$\ }
\def\ap{\alpha'}
\def\Tp{{{T'}^j}}
\def\Up{{{U'}^j}}

\def\oA{{\ov C_{\th}}} \def\oAA{{\ov C_{\nu}}}
\def\A{{C_{\th}}}       \def\AA{{C_{\nu}}}
\def\cf{{\it cf.\ }}
\def\ie{{\it i.e.\ }}
\def\eg{{\it e.g.\ }}
\def\eqq{{\it Eq.\ }}
\def\eqqs{{\it Eqs.\ }}
\def\th{\theta}
\def\Th{\Theta}
\def\eps{\epsilon}
\def\al{\alpha}

\def\si{\sigma}

\def\be{\beta}

\def\Ec{{\cal E}}
\newif\ifnref
\def\rrr#1#2{\relax\ifnref\nref#1{#2}\else\ref#1{#2}\fi}
\def\ldf#1#2{\begingroup\obeylines
\gdef#1{\rrr{#1}{#2}}\endgroup\unskip}
\def\nrf#1{\nreftrue{#1}\nreffalse}

\def\multrefvii#1#2#3#4#5#6#7{\nrf{#1#2#3#4#5#6}\refs{#1{--}#7}}
\def\doubref#1#2{\refs{{#1},{#2} }}
\def\threeref#1#2#3{\refs{{#1},{#2},{#3} }}

\def\Fourref#1#2#3#4{\nrf{#1#2#3#4}\refs{#1{--}#4}}

\nreffalse

\def\lref{\ldf}


\def\appA{A}
\def\appB{B}
\def\appC{C}
\def\tilde{\widetilde}

\def\h {{1\over 2}}

\def\ov {\overline}
\def\o {\over}
\def\fc#1#2{{#1 \o #2}}

\def\IZ{ {\bf Z}}
\def\IC{{\bf C}}
\def\IR{ {\bf R}}
\def\hat{\widehat}

\def\br{\hfill\break}

\def\det {{\rm det}}
\def\mod {{\rm mod}}
\def\lf {\left}
\def\ri {\right}
\def\ra {\rightarrow}
\def\lra {\longrightarrow}
\def\re {{\rm Re}}
\def\im {{\rm Im}}
\def\p {\partial}

\def\Fc {{\cal F}} 
\def\Cc {{\cal C}} \def\Oc {{\cal O}}
 \def\Sc {{\cal S}}
  \def\Ac {{\cal A}}

\lref\GNS{E.~Gava, K.~S.~Narain and M.~H.~Sarmadi,
``On the bound states of p- and (p+2)-branes,''
Nucl.\ Phys.\ B {\bf 504}, 214 (1997)
[arXiv:hep-th/9704006].
}

\lref\Doug{I.~Brunner, M.~R.~Douglas, A.~E.~Lawrence and C.~Romelsberger,
``D-branes on the quintic,''
JHEP {\bf 0008}, 015 (2000)
[arXiv:hep-th/9906200].
}
\lref\KL{V.~Kaplunovsky and J.~Louis,
 ``On Gauge couplings in string theory,''
Nucl.\ Phys.\ B {\bf 444}, 191 (1995)
[arXiv:hep-th/9502077].
}
\lref\ABFPT{
I.~Antoniadis, C.~Bachas, C.~Fabre, H.~Partouche and T.~R.~Taylor,
"Aspects of type I - type II - heterotic triality in four dimensions,''
Nucl.\ Phys.\ B {\bf 489}, 160 (1997)
[arXiv:hep-th/9608012].
}
\lref\all{
R.~Blumenhagen, L.~G\"orlich, B.~K\"ors and D.~L\"ust,
"Noncommutative compactifications of type I strings on tori with  magnetic background flux,''
JHEP {\bf 0010}, 006 (2000)
[arXiv:hep-th/0007024];\br
C.~Angelantonj, I.~Antoniadis, E.~Dudas and A.~Sagnotti,
"Type-I strings on magnetised orbifolds and brane transmutation,''
Phys.\ Lett.\ B {\bf 489}, 223 (2000)
[arXiv:hep-th/0007090];\br
G.~Aldazabal, S.~Franco, L.E.~Ibanez, R.~Rabadan and A.M.~Uranga,
"Intersecting brane worlds,''
JHEP {\bf 0102}, 047 (2001)
[arXiv:hep-ph/0011132]; 
"D = 4 chiral string compactifications from intersecting branes,''
J.\ Math.\ Phys.\  {\bf 42}, 3103 (2001)
[arXiv:hep-th/0011073];\br
R.~Blumenhagen, B.~K\"ors and D.~L\"ust,
"Type I strings with F- and B-flux,''
JHEP {\bf 0102}, 030 (2001)
[arXiv:hep-th/0012156];\br
L.E.~Ibanez, F. Marchesano and R. Rabadan,
"Getting just the standard model at intersecting branes,''
JHEP {\bf 0111}, 002 (2001)
[arXiv:hep-th/0105155];\br
C.~Kokorelis,
"New standard model vacua from intersecting branes,''
JHEP {\bf 0209}, 029 (2002)
[arXiv:hep-th/0205147];\br
R.~Blumenhagen, B.~K\"ors, D.~L\"ust and T.~Ott,
"The standard model from stable intersecting brane world orbifolds,''
Nucl.\ Phys.\ B {\bf 616}, 3 (2001)
[arXiv:hep-th/0107138];\br
M.~Cvetic, G.~Shiu and A.M.~Uranga,
"Chiral four-dimensional N = 1 supersymmetric type $IIA$ orientifolds from  intersecting 
D6-branes,'' Nucl.\ Phys.\ B {\bf 615}, 3 (2001)
[arXiv:hep-th/0107166];\br
C.~Kokorelis,
``Exact standard model structures from intersecting D5-branes,''
Nucl.\ Phys.\ B {\bf 677}, 115 (2004)
[arXiv:hep-th/0207234];\br
D.~Bailin, G.V.~Kraniotis and A.~Love,
"Standard-like models from intersecting D4-branes,''
Phys.\ Lett.\ B {\bf 530}, 202 (2002)
[arXiv:hep-th/0108131];
"New standard-like models from intersecting D4-branes,''
Phys.\ Lett.\ B {\bf 547}, 43 (2002)
[arXiv:hep-th/0208103];\br
R.~Blumenhagen, L.~G\"orlich and T.~Ott,
"Supersymmetric intersecting branes on the type $IIA$ T(6)/Z(4) orientifold,''
JHEP {\bf 0301}, 021 (2003)
[arXiv:hep-th/0211059];\br
R.~Blumenhagen, V.~Braun, B.~K\"ors and D.~L\"ust,
"Orientifolds of K3 and Calabi-Yau manifolds with intersecting $D$--branes,''
JHEP {\bf 0207}, 026 (2002)
[arXiv:hep-th/0206038];\br
C.~Kokorelis,
``Exact standard model structures from intersecting branes,''
arXiv:hep-th/0210004;\br
G.~Honecker,
``Chiral supersymmetric models on an orientifold of Z(4) x Z(2) with  intersecting D6-branes,''
Nucl.\ Phys.\ B {\bf 666}, 175 (2003)
[arXiv:hep-th/0303015];\br
R.~Blumenhagen,
``Supersymmetric orientifolds of Gepner models,''
JHEP {\bf 0311}, 055 (2003)
[arXiv:hep-th/0310244];\br
I.~Brunner, K.~Hori, K.~Hosomichi and J.~Walcher,
``Orientifolds of Gepner models,''
arXiv:hep-th/0401137;\br
R.~Blumenhagen and T.~Weigand,
``Chiral supersymmetric Gepner model orientifolds,''
JHEP {\bf 0402}, 041 (2004)
[arXiv:hep-th/0401148];\br
T.~P.~T.~Dijkstra, L.~R.~Huiszoon and A.~N.~Schellekens,
 ``Chiral supersymmetric standard model spectra from orientifolds of Gepner
models,''
arXiv:hep-th/0403196;\br
G.~Honecker and T.~Ott,
 ``Getting just the supersymmetric standard model at intersecting branes on the
Z(6)-orientifold,''
arXiv:hep-th/0404055;\br
G.~Aldazabal, E.~C.~Andres and J.~E.~Juknevich,
``Particle models from orientifolds at Gepner-orbifold points,''
arXiv:hep-th/0403262.
}

\lref\BlumenhagenTE{
R.~Blumenhagen, B.~K\"ors, D.~L\"ust and T.~Ott,
"The standard model from stable intersecting brane world orbifolds,''
Nucl.\ Phys.\ B {\bf 616}, 3 (2001)
[arXiv:hep-th/0107138].
}

\lref\ACNY{E.S.~Fradkin and A.A.~Tseytlin,
"Nonlinear Electrodynamics From Quantized Strings,''
Phys.\ Lett.\ B {\bf 163}, 123 (1985);\br
A.~Abouelsaood, C.G. Callan, C.R.~Nappi and S.A.~Yost,
"Open Strings In Background Gauge Fields,''
Nucl.\ Phys.\ B {\bf 280}, 599 (1987).
}
\lref\SW{N.~Seiberg and E.~Witten,
"String theory and noncommutative geometry,''
JHEP {\bf 9909}, 032 (1999)
[arXiv:hep-th/9908142].
}

\lref\MSch{J.~Maharana and J.H.~Schwarz,
"Noncompact Symmetries In String Theory,''
Nucl.\ Phys.\ B {\bf 390}, 3 (1993)
[arXiv:hep-th/9207016].
}

\lref\LMS{D.~L\"ust, P. Mayr and S.~Stieberger, work in progress.}

\lref\kona{B.~K\"ors and P.~Nath,
``Effective action and soft supersymmetry breaking for intersecting D-brane
models,''
Nucl.\ Phys.\ B {\bf 681}, 77 (2004)
[arXiv:hep-th/0309167].
}

\lref\CremmerEN{
E.~Cremmer, S.~Ferrara, L.~Girardello and A.~Van Proeyen,
"Yang-Mills Theories With Local Supersymmetry: Lagrangian, Transformation
Nucl.\ Phys.\ B {\bf 212}, 413 (1983).
}

\lref\BL{
M.~Berkooz and R.~G.~Leigh,
"A D = 4 N = 1 orbifold of type I strings,''
Nucl.\ Phys.\ B {\bf 483}, 187 (1997)
[arXiv:hep-th/9605049].
}

\lref\bonn{Z.~Lalak, S.~Lavignac and H.P.~Nilles,
"Target-space duality in heterotic and type I effective Lagrangians,''
Nucl.\ Phys.\ B {\bf 576}, 399 (2000)
[arXiv:hep-th/9912206].
}

\lref\BDL{M.~Berkooz, M.R.~Douglas and R.G.~Leigh,
"Branes intersecting at angles,''
Nucl.\ Phys.\ B {\bf 480}, 265 (1996)
[arXiv:hep-th/9606139].
}

\lref\BGKL{R.~Blumenhagen, L.~G\"orlich, B.~K\"ors and D.~L\"ust,
"Asymmetric orbifolds, noncommutative geometry and type I string vacua,''
Nucl.\ Phys.\ B {\bf 582}, 44 (2000)
[arXiv:hep-th/0003024].
}

\lref\AbelYX{
S.A.~Abel and A.W.~Owen,
"N-point amplitudes in intersecting brane models,''
arXiv:hep-th/0310257.
}

\lref\CremadesCS{
D.~Cremades, L.E.~Ibanez and F.~Marchesano,
"Intersecting brane models of particle physics and the Higgs mechanism,''
JHEP {\bf 0207}, 022 (2002)
[arXiv:hep-th/0203160];
"Towards a theory of quark masses, mixings and CP-violation,''
arXiv:hep-ph/0212064.
}

\lref\bain{P.~Bain and M.~Berg,
"Effective action of matter fields in four-dimensional string  orientifolds,''
JHEP {\bf 0004}, 013 (2000)
[arXiv:hep-th/0003185].
}

\lref\klebanov{I.R.~Klebanov and L.~Thorlacius,
"The Size of p-Branes,''
Phys.\ Lett.\ B {\bf 371}, 51 (1996)
[arXiv:hep-th/9510200].
}

\lref\HK{
A.~Hashimoto and I.R.~Klebanov,
"Decay of Excited $D$--branes,''
Phys.\ Lett.\ B {\bf 381}, 437 (1996)
[arXiv:hep-th/9604065].
}
\lref\HKi{A.~Hashimoto and I.R.~Klebanov,
"Scattering of strings from $D$--branes,''
Nucl.\ Phys.\ Proc.\ Suppl.\  {\bf 55B}, 118 (1997)
[arXiv:hep-th/9611214].
}
\lref\GMi{M.R.~Garousi and R.C.~Myers,
"Superstring Scattering from $D$-Branes,''
Nucl.\ Phys.\ B {\bf 475}, 193 (1996)
[arXiv:hep-th/9603194].
}
\lref\GMii{
M.R.~Garousi and R.C.~Myers,
"World-volume interactions on $D$--branes,''
Nucl.\ Phys.\ B {\bf 542}, 73 (1999)
[arXiv:hep-th/9809100].
}
\lref\GMiii{M.R.~Garousi and R.C.~Myers,
"World-volume potentials on $D$--branes,''
JHEP {\bf 0011}, 032 (2000)
[arXiv:hep-th/0010122].
}

\lref\AKI{A.~Hashimoto,
"Dynamics of Dirichlet-Neumann open strings on D-branes,''
Nucl.\ Phys.\ B {\bf 496}, 243 (1997)
[arXiv:hep-th/9608127].
}

\lref\benakli{I.~Antoniadis, K.~Benakli and A.~Laugier,
"Contact interactions in $D$--brane models,''
JHEP {\bf 0105}, 044 (2001)
[arXiv:hep-th/0011281].
}

\lref\GrossDD{
D.J.~Gross, J.A.~Harvey, E.J.~Martinec and R.~Rohm,
"The Heterotic String,''
Phys.\ Rev.\ Lett.\  {\bf 54}, 502 (1985).
}

\lref\BachasIK{
C.~Bachas,
"A Way to break supersymmetry,''
arXiv:hep-th/9503030.
}

\lref\KiritsisMC{
E.~Kiritsis,
"$D$--branes in standard model building, gravity and cosmology,''
arXiv:hep-th/0310001;\br
D. L\"ust,
"Intersecting brane worlds: A path to the standard model?,''
arXiv:hep-th/0401156.
}

\lref\LauerKH{D.~L\"ust, S.~Theisen and G.~Zoupanos,
"Four-Dimensional Heterotic Strings And Conformal Field Theory,''
Nucl.\ Phys.\ B {\bf 296}, 800 (1988);\br
J.~Lauer, D.~L\"ust and S.~Theisen,
"Four-Dimensional Supergravity From Four-Dimensional Strings,''
Nucl.\ Phys.\ B {\bf 304}, 236 (1988).
}

\lref\DKL{
L.J.~Dixon, V.~Kaplunovsky and J.~Louis,
"On Effective Field Theories Describing (2,2) Vacua Of The Heterotic String,''
Nucl.\ Phys.\ B {\bf 329}, 27 (1990).
}

\lref\KaplunovskyRD{
V.S.~Kaplunovsky and J.~Louis,
"Model independent analysis of soft terms in effective supergravity and in string theory,''
Phys.\ Lett.\ B {\bf 306}, 269 (1993)
[arXiv:hep-th/9303040];\br
A.~Brignole, L.E.~Ibanez and C.~Munoz,
"Towards a theory of soft terms for the supersymmetric Standard Model,''
Nucl.\ Phys.\ B {\bf 422}, 125 (1994)
[Erratum-ibid.\ B {\bf 436}, 747 (1995)]
[arXiv:hep-ph/9308271].
}

\lref\BlumenhagenJY{
R.~Blumenhagen, D.~L\"ust and S.~Stieberger,
"Gauge unification in supersymmetric intersecting brane worlds,''
JHEP {\bf 0307}, 036 (2003)
[arXiv:hep-th/0305146].
}

\lref\AntoniadisEN{
I.~Antoniadis, E.~Kiritsis and T.N.~Tomaras,
"A $D$--brane alternative to unification,''
Phys.\ Lett.\ B {\bf 486}, 186 (2000)
[arXiv:hep-ph/0004214].
}

\lref\IbanezHC{
L.E.~Ibanez and D.~L\"ust,
"Duality anomaly cancellation, minimal string unification and the effective low-energy Lagrangian 
of 4-D strings,''
Nucl.\ Phys.\ B {\bf 382}, 305 (1992)
[arXiv:hep-th/9202046];\br
H.P.~Nilles and S.~Stieberger,
"String unification, universal one-loop corrections and strongly coupled  heterotic string 
theory,'' 
Nucl.\ Phys.\ B {\bf 499}, 3 (1997)
[arXiv:hep-th/9702110].
}

\lref\DFMS{L.J.~Dixon, D.~Friedan, E.J.~Martinec and S.H.~Shenker,
"The Conformal Field Theory Of Orbifolds,''
Nucl.\ Phys.\ B {\bf 282}, 13 (1987).
}

\lref\HV{S.~Hamidi and C.~Vafa,
"Interactions On Orbifolds,''
Nucl.\ Phys.\ B {\bf 279}, 465 (1987).
}

\lref\KW{I.R.~Klebanov and E.~Witten,
"Proton decay in intersecting $D$--brane models,''
Nucl.\ Phys.\ B {\bf 664}, 3 (2003)
[arXiv:hep-th/0304079].
}

\lref\LS{D.~L\"ust and S.~Stieberger,
"Gauge threshold corrections in intersecting brane world models,''
arXiv:hep-th/0302221.
}

\lref\david{J.R.~David,
``Tachyon condensation in the D0/D4 system,''
JHEP {\bf 0010}, 004 (2000)
[arXiv:hep-th/0007235];
"Tachyon condensation using the disk partition function,''
JHEP {\bf 0107}, 009 (2001)
[arXiv:hep-th/0012089].
}

\lref\abel{S.A.~Abel and A.W.~Owen,
"Interactions in intersecting brane models,''
Nucl.\ Phys.\ B {\bf 663}, 197 (2003)
[arXiv:hep-th/0303124].
}

\lref\uranga{M.~Cvetic, G.~Shiu and A.M.~Uranga,
"Chiral four-dimensional N = 1 supersymmetric type $IIA$ orientifolds from  
intersecting D6-branes,'' Nucl.\ Phys.\ B {\bf 615}, 3 (2001)
[arXiv:hep-th/0107166];\br
M.~Cvetic, I.~Papadimitriou and G.~Shiu,
"Supersymmetric three family SU(5) grand unified models from type $IIA$  orientifolds with 
arXiv:hep-th/0212177;\br
M.~Cvetic and I.~Papadimitriou,
"More supersymmetric standard-like models from intersecting D6-branes on  
Phys.\ Rev.\ D {\bf 67}, 126006 (2003)
[arXiv:hep-th/0303197].
}

\lref\cvetic{M.~Cvetic and I.~Papadimitriou,
"Conformal field theory couplings for intersecting $D$--branes on  orientifolds,''
Phys.\ Rev.\ D {\bf 68}, 046001 (2003)
[arXiv:hep-th/0303083].
}

\lref\CIM{D.~Cremades, L.E.~Ibanez and F.~Marchesano,
"Yukawa couplings in intersecting $D$--brane models,''
JHEP {\bf 0307}, 038 (2003)
[arXiv:hep-th/0302105].
}

\lref\DKL{L.J.~Dixon, V.~Kaplunovsky and J.~Louis,
"On Effective Field Theories Describing (2,2) Vacua Of The Heterotic String,''
Nucl.\ Phys.\ B {\bf 329}, 27 (1990).
}

\lref\EJSS{
S.~Stieberger, D.~Jungnickel, J.~Lauer and M.~Spalinski,
"Yukawa couplings for bosonic ${\bf Z}_N$ orbifolds: 
Mod.\ Phys.\ Lett.\ A {\bf 7}, 3059 (1992)
[arXiv:hep-th/9204037];\br
J.~Erler, D.~Jungnickel, M.~Spalinski and S.~Stieberger,
"Higher twisted sector couplings of ${\bf Z}_N$ orbifolds,''
Nucl.\ Phys.\ B {\bf 397}, 379 (1993)
[arXiv:hep-th/9207049];\br
S.~Stieberger,
"Moduli and twisted sector dependence on ${\bf Z}_N\times {\bf Z}_M$ orbifold couplings,''
Phys.\ Lett.\ B {\bf 300}, 347 (1993)
[arXiv:hep-th/9211027].
}

\lref\BKM{T.T.~Burwick, R.K.~Kaiser and H.F.~Muller,
"General Yukawa Couplings Of Strings On ${\bf Z}_N$ Orbifolds,''
Nucl.\ Phys.\ B {\bf 355}, 689 (1991).
}
\lref\casas{J.A.~Casas, F.~Gomez and C.~Munoz,
"Complete structure of ${\bf Z}_N$ Yukawa couplings,''
Int.\ J.\ Mod.\ Phys.\ A {\bf 8}, 455 (1993)
[arXiv:hep-th/9110060].
}

\lref\koba{P.~Mayr and S.~Stieberger,
"Low-energy properties of (0,2) compactifications,''
arXiv:hep-th/9412196;\br
T.~Kobayashi and O.~Lebedev,
"Heterotic Yukawa couplings and continuous Wilson lines,''
Phys.\ Lett.\ B {\bf 566}, 164 (2003)
[arXiv:hep-th/0303009].
}

\lref\gauge{G.~Shiu and S.H.~Tye,
"TeV scale superstring and extra dimensions,''
Phys.\ Rev.\ D {\bf 58}, 106007 (1998)
[arXiv:hep-th/9805157];\br
D.~Cremades, L.E.~Ibanez and F.~Marchesano,
"SUSY quivers, intersecting branes and the modest hierarchy problem,''
JHEP {\bf 0207}, 009 (2002)
[arXiv:hep-th/0201205];\br
M.~Cvetic, P.~Langacker and G.~Shiu,
"Phenomenology of a three-family standard-like string model,''
Phys.\ Rev.\ D {\bf 66}, 066004 (2002)
[arXiv:hep-ph/0205252].
}

\lref\IBA{D.~Cremades, L.E.~Ibanez and F.~Marchesano,
"SUSY quivers, intersecting branes and the modest hierarchy problem,''
JHEP {\bf 0207}, 009 (2002)
[arXiv:hep-th/0201205].}

\lref\FLT{S.~Ferrara, D.~L\"ust, A.D.~Shapere and S.~Theisen,
"Modular Invariance In Supersymmetric Field Theories,''
Phys.\ Lett.\ B {\bf 225}, 363 (1989);\br
S.~Ferrara, D.~L\"ust and S.~Theisen,
"Target Space Modular Invariance And Low-Energy Couplings In Orbifold Compactifications,''
Phys.\ Lett.\ B {\bf 233}, 147 (1989).
}

\lref\LMN{J.~Lauer, J.~Mas and H.P.~Nilles,
"Duality And The Role Of Nonperturbative Effects On The World Sheet,''
Phys.\ Lett.\ B {\bf 226}, 251 (1989);
"Twisted Sector Representations Of Discrete Background Symmetries For Two-Dimensional Orbifolds,''
Nucl.\ Phys.\ B {\bf 351}, 353 (1991).
}

\lref\robert{R. Richter, {\it Scattering of moduli and matter fields
in type $II$  superstring backgrounds}, Diploma thesis, December 2003.}

\lref\AFIV{G.~Aldazabal, A.~Font, L.E.~Ibanez and G.~Violero,
"D = 4, N = 1, type IIB orientifolds,''
Nucl.\ Phys.\ B {\bf 536}, 29 (1998)
[arXiv:hep-th/9804026];\br
L.E.~Ibanez, C.~Munoz and S.~Rigolin,
"Aspects of type I string phenomenology,''
Nucl.\ Phys.\ B {\bf 553}, 43 (1999)
[arXiv:hep-ph/9812397].
}

\lref\MSD{
P.~Mayr and S.~Stieberger,
``Dilaton, antisymmetric tensor and gauge fields in string effective theories
at the one loop level,''
Nucl.\ Phys.\ B {\bf 412}, 502 (1994)
[arXiv:hep-th/9304055].
}

\lref\joep{
J.~Polchinski, S.~Chaudhuri and C.V.~Johnson,
"Notes on D-Branes,''
arXiv:hep-th/9602052;\br
J.~Polchinski, 
"String Theory'', Vol. 2, Section 13, Cambridge University Press 1998.}

\lref\IRAN{F.~Ardalan, H.~Arfaei and M.M.~Sheikh-Jabbari,
"Noncommutative geometry from strings and branes,''
JHEP {\bf 9902}, 016 (1999)
[arXiv:hep-th/9810072].
}

\Title{\vbox{\rightline{HU--EP--03/83} 
\rightline{MPP--2003--145}
\rightline{LMU--TPW 02/04}
\rightline{\tt hep-th/0404134} \vskip-2cm }}
{\vbox{\centerline{Scattering of Gauge, Matter, and Moduli Fields}
\bigskip\centerline{from Intersecting Branes}}}
\vskip-7pt
\centerline{D. L\"ust$^{a,b}$,\  P. Mayr$^c$,\
R. Richter$^a$, and\ S. Stieberger$^a$}
\bigskip
\centerline{\it $^a$ Institut f\"ur Physik, Humboldt Universit\"at zu Berlin,}
\centerline{\it Newtonstra\ss e 15, 12489 Berlin, FRG}
\vskip4pt
\centerline{\it $^b$ Max--Planck--Institut f\"ur Physik,}
\centerline{\it F\"ohringer Ring 6, 80805 M\"unchen, FRG}
\vskip4pt
\centerline{\it $^c$ Sektion Physik, 
Ludwig-Maximilians-Universit\"at M\"unchen,}
\centerline{\it Theresienstra\ss e 37, 80333 M\"unchen, FRG}

\bigskip\bigskip
\centerline{\bf Abstract}
\vskip .2in
\noindent

We calculate various tree--level (disk) 
scattering amplitudes involving gauge, matter and moduli fields in \tb toroidal 
orbifold/orientifold backgrounds with $D9,D5$ respectively $D7,D3$--branes 
or via $T$-duality $D6$--branes in type $IIA$ compactifications.
In type $IIB$ the $D$--branes  
may have non--vanishing fluxes on their  world--volume. From 
these results we extract the moduli and flux dependence of the 
tree--level gauge couplings, the metrics for the moduli and matter fields.
The non-vanishing fluxes correspond in the $T$--dual type $IIA$ description 
to intersecting $D6$--branes.
This allows us to determine the moduli dependence of the tree--level matter field metrics 
in the effective action of intersecting $D6$--brane models. In addition we
derive the physical Yukawa couplings with their correct normalization.

\Date{}
\noindent

\goodbreak

\newsec{Introduction}

Four-dimensional superstrings constitute at the moment the
best candidates for unification of all interactions.
In trying to use these theories to describe the observed physics
or the measurements in future accelerator experiments
like the LHC, it is of fundamental importance to obtain the
low-energy field theory of each given class of 4-D string.
In  this context the deviation of the minimal supersymmetric
standard model (MSSM), 
the issue of supersymmetry breaking, mass generation,
the search
for supersymmetric particles, rare decays, flavor changing neutral currents,
the search for extra dimensions and
gauge coupling unifications will be central topics.

About ten years ago this program was pursued with great detail
and also some success (see e.g  \Fourref\LauerKH\DKL\IbanezHC\KaplunovskyRD) 
for 4-D heterotic string vacua \GrossDD .
More recently it became evident that
open string constructions and most notably 
intersecting brane worlds 
\threeref\BachasIK\BDL\all\ offer excellent opportunities for realistic string phenomenology
(see also \KiritsisMC\ for a review).
Within these string compactifications,  in general
a hierarchy of dimensions supporting different kinds of massless fields is
opening up: 
first, there is the closed string sector in the entire ten--dimensional space,
which contains besides the gravitational
fields also the geometric scalar moduli fields of the internal compact space.
Second, open strings provide the
standard model gauge fields,
living on the $p+1$--dimensional world volumes of $Dp$--branes, and third,
the standard model matter fields correspond
to open string excitations which are located at the even
lower-dimensional various intersections of the $D$--branes in the internal six--dimensional space. 
Then the fermion spectrum is determined by 
the intersection numbers of certain homology--cycles in the internal 
space, as opposed to the heterotic 
strings, where the number of generations was given by the Euler characteristic 
in the simplest case.

Some results on the low-energy effective action of intersecting
brane world models were already obtained in the past:
the effective
scalar potential is needed to discuss the question of stability
of intersecting brane world models \BlumenhagenTE;
tree-level gauge couplings \gauge\ and one-loop gauge threshold corrections \LS\
in supersymmetric intersecting brane world models were calculated,
and the question of gauge coupling unification
was addressed in \doubref\AntoniadisEN\BlumenhagenJY;
using the N=1 \ti-- heterotic string--string duality,
some results for the matter field K\"ahler metric and for soft-SUSY
breaking terms in the effective action were obtained in \kona;
finally effective Yukawa couplings 
\threeref\CremadesCS\CIM\cvetic\ and higher point scattering  \AbelYX   ,
relevant for flavor changing neutral currents \abel\
and proton decay \KW, were also investigated. 
However, the amplitudes discussed in all those references involve only open strings 
inserted at the boundary. In certain cases the amplitude may be easily deduced from known
closed string results. More concretely, the open string vertex operators inserted at the boundary 
depend only holomorphically on the world--sheet coordinates. If the full open
string vertex operator or only its internal part represents just half of an
analogous closed string vertex operator, for that particular piece of the full correlator
a similar computation on the sphere may be borrowed. 
The latter involves both the holomorphic and anti--holomorphic side and the open string
result may be obtained  by essentially ``taking the square root''. 
However already in the case
of scalar matter field vertex operators, this simple procedure fails. E.g. a scattering of 
four matter fields on the sphere like in heterotic string theory 
assumes a completely different momentum expansion than the same amplitude with matter fields
represented by open strings. To conclude, generically one should not expect a closed string 
computation in any simple way to be related to a scattering amplitude involving open string
states from the boundary. As we shall see, this statement will be strengthened when 
considering disk amplitudes involving in addition also closed string states
from the bulk.
These bulk states in particular include all closed string moduli of the
underlying compact fields, namely the K\"ahler moduli $T^j$ as well as the
complex structure moduli $U^j$. Therefore, deriving the moduli dependent gauge
coupling constants for the open string gauge fields or the moduli dependent K\"ahler
metrics for the open string matter fields, mixed open/closed string scattering
amplitudes have to be computed. As we will see, here subtle issues
arise, which are quite different from heterotic string amplitudes.

The aim of this paper is to explicitly compute several disk 
scattering amplitudes involving
open string gauge and  matter fields as well as closed
string moduli fields
from intersecting branes and deduce from them directly relevant parts 
of the open/closed string 4-D effective, low-energy field theory. 
We will perform the computation of the open/closed string scattering
amplitudes in the type $IIB$ ($\Fc$-flux picture) as well as in the $T$--dual type $IIA$
picture (angle picture).
In type $IIB$ we are dealing with $D9$--branes wrapped
on a six-dimensional torus with open string magnetic $\Fc$--flux turned on.
Therefore the $D9$-branes have in general mixed Dirichlet and Neumann
boundary conditions on the torus, which becomes non-commutative in this way.
Via $T$--duality on three directions of the torus the $D9$--branes
become $D6$--branes, being wrapped around 3--cycles of the
dual torus and intersecting each other on various
points of the dual torus, with intersection angles
that are in one-to-one correspondences with the $IIB$ $\Fc$--fluxes.
All our disk amplitudes are valid for arbitrary flux values, i.e.
arbitrary intersection angles. However when extracting from the
amplitudes the low-energy effective action we mainly concentrate
on $D$--brane configurations which preserve 4-D 
N=1 space-time supersymmetry. In type $IIB$ this means that
the fluxes have to be chosen in such a way that essentially the
$D$--branes have  either $D9$-- or $D5$--brane (or $D7/D3$--brane) boundary conditions.
On the $T$--dual other side, the $D6$--branes have to satisfy
certain angle conditions in order to be N=1 space-time
supersymmetric.

It is well known \CremmerEN\ that any N=1 supergravity action in four space--time dimensions
is encoded by three functions, namely the K\"ahler potential $K$,
the superpotential $W$, and the gauge kinetic function $f$.
When such an effective action arises from a higher dimensional string theory 
these three function usually depend (non--trivially) 
on moduli fields describing 
the background of the present string model.
It is the aim of this article to determine the moduli 
dependence of the gauge kinetic function $f$ and of the K\"ahler potential $K$
from 
string tree--level scattering amplitudes.
One--loop corrections will be discussed elsewhere.

In the first part of the paper we will compute those scattering
amplitudes between open string gauge bosons and 
closed string moduli fields that are relevant
for the moduli-dependent, effective 4-D gauge couplings.
This is important for the issue of gauge coupling unification and, in
supersymmetric compactifications, for the computation of the
soft gaugino masses.
The holomorphic gauge kinetic function has been 
determined for supersymmetric intersecting brane 
world models,  in \gauge\ at string  tree--level via dimensional
reduction of the Born-Infeld action and in \LS\ at one--loop by explicit string calculations. 
Our results indeed here confirm the findings of \gauge\ for supersymmetric
$D$--brane configurations.

Secondly, we compute the kinetic energy terms in the effective
action for open string matter fields. Being moduli--dependent
quantities, these are needed for
the correct normalization of the (physical) Yukawa couplings, for the matter field
scalar potential and  for the soft scalar masses (squarks and sleptons). 
In intersecting brane world models chiral fermions 
stem from open strings stretched
between two intersecting branes of angles $\th=(\th^1,\th^2,\th^3)$.
For supersymmetric angles,
these fermions come in N=1 chiral multiplets, whose lowest 
component is the scalar matter field $C_\th$. From 
the point of view of the open string conformal field theory,
the open fields $C_\th$ obey 
mixed boundary conditions and are associated to twist fields.
In addition there may be untwisted matter fields $C_i$, which
come in N=4 multiplets for the case of toroidal models.

The K\"ahler 
potential for the moduli $M$ and matter fields $C_i,C_\th$ reads up to second order 
in the matter fields: 
\eqn\kaehler{
K=\hat{K}(M,\ov M)+\sum_{twisted\atop matter\ \th} G_{C_\th\ov C_\th}(M,\ov M) 
\ C_\th\ov C_\th+\sum_{untwisted\atop matter\ i,j} G_{C_i\ov C_j}(M,\ov M)\ 
C_i\ov C_j+\Oc(C^4)\ .}
The index  $\th$ collectively denotes a given matter field coming from one
pair of intersecting branes of angles $\th$. 
It is the purpose of this article to calculate the metrics 
$G_{C_\th\ov C_\th}$ and
$G_{C_i\ov C_j}$ for intersecting brane worlds
from various disk amplitudes involving matter and moduli fields.
Untwisted matter field metrics for N=1 type $I$ compactifications on toroidal orbifolds 
with parallel $D9$--branes without fluxes have been determined in \bain.
In addition we will also calculate open string disk amplitudes that
describe the scattering of four 
matter fields. Performing suitable factorization
limits these amplitudes contain useful informations on the physical
Yukawa couplings and in this way on the (holomorphic)
trilinear superpotential which
reads:
\eqn\superpot{
W=\sum_{\al,\be,\gamma} W_{\al\be\gamma}(t^l)\ C_\al C_\be C_\gamma \ .}

Our string scattering computations for the matter field metrics substantially extend
and improve the claims made in \kona.
The latter relies on arguments based on N=1 type $I$ -- heterotic duality in $D=4$.
This is a strong--weak coupling duality involving a 
mixing of the dilaton with the other moduli fields. 
Therefore a comparison of heterotic results on the sphere with a similar
\ti result from the disk is to be questioned. Moreover, we shall see that 
the moduli dependence is derived from a disk amplitude
involving both open strings from the brane and closed strings from the bulk. 
This coupling
has no obvious relation to a pure closed string coupling on the sphere. 
In particular, this disk coupling is not just a  square root of a similar 
coupling on the sphere.
Furthermore, taking into account the problems discussed in \bonn, 
it is a challenging 
task to find the correct duality map of fields in the effective actions of
N=1 heterotic and \ti theories.

The paper is organized as follows. In the next section we review
the aspects of open string conformal field theory which will become
relevant for the computation of the various disk amplitudes. Special
emphasis is given to $D$--branes with mixed boundary conditions (type
$IIB$ picture), respectively intersecting $D$--branes (type $IIA$).
Section 3 is devoted to the derivation of the tree level effective
gauge couplings. First, we compute the gauge couplings from the
Born-Infeld action with fluxes in type $IIB$,  including also the known expressions \gauge\ of
the holomorphic $f$-function for supersymmetric $D9$-- and $D5$--branes without fluxes.
These results are nicely reproduced by the scattering
of two open string gauge fields with one closed string modulus,
where one has to make a careful transformation between the string moduli, referring to 
string vertex operators,  and the (supergravity) field theory moduli fields.
Analogous results are obtained for rotated $D6$--branes in the $T$--dual
type $IIA$ picture.
In sections 4 and 5 we present our results on the scattering of
two open string matter fields and one or two closed string moduli.
First, in section 4, we derive the amplitudes and the metric for the untwisted
matter fields in the presence of fluxes on the branes. After that in section 5
we discuss the twist field
open string vertex operators and the 
associated disk amplitudes and compute the twist
field matter metrics.
Finally in section 6 we compute the scattering amplitudes of four
open string matter fields which are needed for the derivation
of the physical Yukawa couplings. Some conclusions
are presented in section 7.
In appendix \appA\ we consider scattering of two closed string moduli fields off
$Dp$--branes. These results allow to determine the K\"ahler potential for the closed string moduli
fields. In appendix \appB\ we calculate disk amplitudes involving two matter fields from
the boundary and one massless bosonic closed $NS$--string state from the supergravity multiplet.


\newsec{Disk amplitudes involving open and closed string states}

In this subsection we review some technical details occurring in open string
disk calculations.
The world--sheet diagram of a string $S$--matrix describing the interaction of 
open and closed strings at (open string) tree--level can be conformally mapped to a surface with 
one boundary. The latter may be described by a disk, which is conformally 
equivalent to the upper (complex) half--plane ${\cal H}_+=\{z\in \IC\ |\ \im(z)>0\}$. 
The open string states are inserted by vertex operators 
at the boundary of the disk. On the other hand, the closed string vertex operators are 
inserted inside the disk. In theories with $D$--branes massless fields 
like gauge fields, Wilson line moduli or matter fields originate from open string excitations 
living on the $D$--brane world--volume.
Hence the boundary of the disk diagram is attached to the $D$--brane world--volume.
On the other hand, the closed strings, representing \eg the graviton, dilaton and metric moduli 
live in the bulk and are inserted in the bulk of the disk.
Disk scattering of closed massless strings from the supergravity multiplet 
in the presence of $D$--branes has been addressed in the past in Refs. 
\multrefvii\klebanov\HK\HKi\GMi\AKI\GMii\GMiii.
It is one of the purposes of this article to extend these works to moduli fields.

The disk, represented as upper half plane ${\cal H}_+$, may be obtained from the full complex plane
representing the sphere, through a $\IZ_2$ projection $z\mapsto \ov z$.
Apart from the usual  correlators on the sphere 
\eqn\green{\eqalign{
\vev{X^a(z_1)\ X^b(z_2)}&=-g^{ab}\ \ln(z_1-z_2)\ ,\cr
\vev{\psi^{a}(z_1)\ \psi^{b}(z_2)}&=\fc{g^{ab}}{z_1-z_2}\ ,\cr
\vev{\phi(z_1)\ \phi(z_2)}&=-\ln(z_1-z_2)}}
for the bosonic fields $X^a$, fermionic fields $\psi^a$ and ghost field $\phi$,
this projection implies an interaction between the left--moving and right moving 
closed string fields (\cf \eg \GMi):
\eqn\greeni{\eqalign{
\vev{X^a(z_1)\ \tilde X^b(\ov z_2)}&=-D^{ab}\ \ln(z_1-\ov z_2)\ ,\cr
\vev{\psi^a(z_1)\ \tilde \psi^{b}(\ov z_2)}&=\fc{D^{ab}}{z_1-\ov z_2}\ ,\cr
\vev{\phi(z_1)\ \tilde \phi(\ov z_2)}&=-\ln(z_1-\ov z_2)\ .}}
The matrix $D^{ab}$ depends on whether Dirichlet or Neumann boundary conditions are imposed
on the open string fields attached to the  $Dp$--brane:
\eqn\dmatrix{
D^{ab}=\cases{ g^{ab}\ \ ,\ \ &{\rm tangent indices}\ ,\cr
                  -g^{ab}\ \ ,\ \ &{\rm normal indices}\ .}}

We shall calculate disk correlators involving $N_o$ open strings inserted at the boundary $z=\ov z$
of the disk and $N_c$ closed strings from the bulk.
More precisely, we concentrate on the following 
kinds of disk amplitudes $(i)\ N_o=4\,\ N_c=0$,
$(ii)\ N_o=2\ ,\ N_c=1$, $(iii)\ N_o=0\ ,\ N_c=2$ and $(iv)\ N_o=2\ ,\ N_c=2$.
The case $(i)$ is much easier to handle than the other cases.
In the case $(i)$ only massless open string states are inserted at the boundary of ${\cal H}_+$.
The latter describe massless fields like gauge fields or Wilson lines
living on the $D$--brane world volume. Their vertex operators involve only holomorphic fields 
and \eqq \green\ describe their interaction. The same holds
for charged matter fields living on the $D$--brane world volume (or intersections thereof).
On the other hand, if $N_c\neq 0$
also anti--holomorphic fields are involved and produce non--trivial interactions \greeni\ 
between holomorphic and anti--holomorphic fields.
This additional mixing  may simply be taken into account with the so--called 
"doubling trick''. All left--moving fields of the closed string vertex operator 
are multiplied by the matrix $D$:
\eqn\substitute{
\tilde X^i(\ov z)\ra D^i_{\ j}\ X^j(\ov z)\ \ \ ,\ \ \ 
\tilde \psi^i(\ov z)\ra D^i_{\ j}\ \psi^j(\ov z)\ \ \ ,\ \ \ 
\tilde\phi(\ov z)\ra\phi(\ov z)\ .}
Their interactions \greeni\ then follow from the sphere correlators \green\ after taking 
into account the matrix $D^{ij}$ in \substitute.
Essentially this means that the world--sheet is doubled to a sphere with left-- and right--moving
parts of one closed string vertex (inserted at the point $z\in {\cal H}_+$) 
with momentum 
\eqn\closedm{
q^{||}:=\h(q+Dq)} 
assumed to constitute two open string vertex operators. One of the open strings is inserted with 
momentum $q/2$ on the sphere at $z$  and the other with momentum $Dq/2$ at the location $\ov z$.
Hence a disk scattering of $N_o$ open strings inserted at the boundary $z=\ov z$
and $N_c$ closed strings from the bulk is similar to a string scattering of 
$N_o+2N_c$ open strings on the double cover. There are also differences,
namely the Chan-Paton factors, the  
position on the world-sheet and extra constraints on the momenta and polarizations.

Let us stick to our case $(ii)$ involving two open strings of momenta $p_1$ and $p_2$, respectively
and one closed string of momentum $q$. Only the momentum along the direction of the
$D$--brane is conserved:
\eqn\mconserve{p_1+p_2+q^{||}=0\ .}
We may write the relevant closed string momentum $q^{||}=\h(q+Dq)$. According
to the discussion above, we may split
the closed string into two open strings. Therefore, we
have four open strings with momenta $p_1,p_2,\h p$ and $\h Dp$, respectively.
Note, that in the four open string vertex operators the doubled momenta $k_i$ enter:
\eqn\momentum{
k_1=2p_1\ ,\  k_2=2p_2\ ,\ k_3=q\ ,\ k_4=Dq\ .}
With these momenta $k_i$ the momentum conservation \mconserve\ may be written:
\eqn\conserve{
k_1+k_2+k_3+k_4=0}
like in an usual four particle scattering process. Since we consider massless strings only, 
\ie $p_1^2=p_2^2=q^2=0$, we have $k_i^2=0$.
We also introduce the kinematic invariants ($s+t+u=0$):
\eqn\mandelstamm{
s=k_1k_2\ \ \ ,\ \ \ t=k_1k_3\ \ \ ,\ \ \ u=k_1k_4\ .}
However, as already anticipated above, all four momenta are not independent.
This has the consequence \HKi:
\eqn\cons{
s=-2t\ \ \ ,\ \ \ u=t\ ,}
which means, that a scattering of two open strings with one closed string allows only 
for one independent kinematic variable.
We shall always work with the momenta $k_i$, \ie we use $k_i$ in the open string vertex
operator, while the closed string vertex carries the momentum $q$.

So far we have only discussed the case of pure Neumann or Dirichlet boundary
conditions. These two cases are encoded in the matrix $D$, introduced in \dmatrix.
Quite generally, including the case of mixed boundary conditions, which are 
governed by a non--trivial background flux $\Fc$ on the $D$--brane, the matrix $D$ is given by:
\eqn\Dmatrix{
D=-g^{-1}+2\ (g+\Fc)^{-1}\ ,}
as follows by comparing the correlators \green\ and \greeni\ with 
the analogous expressions in \doubref\ACNY\SW.
The matrix $D$ captures the special cases $\Fc=0$ and $\Fc\ra\infty$ corresponding
to pure Neumann or Dirichlet boundary conditions, respectively. For these two cases the
matrix $D$ boils down to \dmatrix.
With the open string metric $G$ and the non--commutativity parameter $\Theta$
\eqn\witten{\eqalign{
\Theta^{-1}&=-(g+\Fc)^{-1}\ \Fc\ (g-\Fc)^{-1}\ ,\cr
G^{-1}&=(g+\Fc)^{-1}\ g\ (g-\Fc)^{-1}\ ,}}
introduced\foot{Compared to \SW, we performed the replacement: $2\pi\ap B\ra \Fc$ and 
$\th^{-1}\ra 2\pi\ap\ \Th^{-1}$.}  
in \SW, we may also write\foot{Note: $D_{\rm symm}=\h(D+D^t)=-g^{-1}+2\ G^{-1}$ and 
$D_{\rm anti}=\h(D-D^t)=2\ \Theta^{-1}$.}:
\eqn\alsoDmatrix{
D=-g^{-1}+2\ G^{-1}+2\ \Theta^{-1}\ .}

\newsec{Tree--level scattering of two gauge fields and one modulus}

In this section we shall investigate quite generally the tree--level gauge couplings of 
$Dp$--branes wrapped on $p-3$--cycles ($p\geq 3$).
As we shall demonstrate in subsection 3.1 these couplings may be easily obtained by dimensionally 
reducing the Born--Infeld action of the $Dp$--brane on the $p-3$--cycle $\Cc_{p-3}$.
But we consider it to be a useful warm up to derive this result independently 
by a string scattering amplitude in order to introduce some crucial ingredients necessary
for the determination of the matter field metric.
We shall determine the tree--level gauge couplings 
of $Dp$--branes with fluxes expressed by the complex moduli fields describing the internal
compactification manifold. This generalizes the discussions in Refs. \doubref\AFIV\gauge.
In particular our formulae capture also intersecting branes with tilted tori.
After performing $T$--duality in $n$ directions, we generically
obtain  the gauge couplings on  tilted $D(p-n)$--branes. 
We set up the discussion in type $IIB$. Hence we consider $D9, D7, D5$ and $D3$--branes
with a $U(N)$ gauge theory on their world--volume.
The  $D9, D7, D5$--branes are wrapped on 
$6,4,2$--cycles, respectively. In addition, we assume the cycles to
be factorizable into $2$--cycles $T^{2,j}\ ,j=1,2,3$. 
The discussion takes over to type $I$ after dropping the $D7$ and $D3$--brane. 
The \ti theory may be obtained from \tb by an orientifold projection.
The world--volume gauge theory on the $D$--brane sitting on the orientifold plane
becomes then $SO(2N)$ or $USp(N)$.
In that case, all gauge couplings, derived in the following for $U(N)$ gauge groups, 
have to be multiplied by a factor of $2$, \ie $g^2_{Dp,SO(2N)}=2g_{Dp}^2$ \joep. 

\subsec{Gauge couplings of $Dp$--branes with fluxes (Type $IIB$ picture)}

We start with the \tb superstring in ten space--time dimensions ($D=10$). 
Its Einstein term is given by
\eqn\Einstein{
\Sc_{IIB}=-\fc{1}{(2\pi)^7\ \alpha'^4}\ \int d^{10}x\ \sqrt{-g_{10}}\ e^{-2\phi_{10}}\  R\ ,}
with the dilaton field $\phi_{10}$ in $D=10$. 
On the other hand the dynamics of the massless $NS$ fields on the $Dp$--brane
are described by the Born--Infeld action
\eqn\BI{
\Sc_{p}=-T_p\ \int d^{p+1}\xi\ e^{-\phi_{10}}\ \Tr\ \sqrt{-\det(G+B+2\pi\ap\ F)}\ ,}
with the world volume gauge field strength $F$. 
The metric $G$ and the anti--symmetric tensor $B$ are the pull--backs of the bulk tensors
to the $D$--brane world--volume.
The $Dp$--brane tension, describing the coupling of the 
gauge boson on the brane to the closed string dilaton in the bulk, is given by 
$T_p=(2\pi)^{-p}\ap^{-\h-\fc{p}{2}}$ (see \eg \joep\ for details).
We compactify \tb on a six--torus $T^6$. The latter is assumed to be a direct
product of three
two--tori, \ie $T^6=\otimes_{j=1}^3 T^{2,j}$. Their corresponding metrics $g^j$
\eqn\metrics{
g^j=\pmatrix{(R^j_1)^2&R^j_1R^j_2\cos\al^j\cr
R^j_1R^j_2\cos\al^j&(R_2^j)^2}} 
lead to the following complexified K\"ahler and complex structure moduli:
\eqn\stringmoduli{
T^j=b^j+i R_1^jR_2^j\sin\al^j\ \ ,\ \ U^j=\fc{R_2^j}{R_1^j}\ e^{i\al^j}\ \ \ ,\ \ \ j=1,2,3\ .}
Here $b^j$ is the value internal NS $B$-field, which i.g. is restricted by the 
projection and not a dynamical scalar in the theory. As discussed below, in type $I$
the real part of the physical field is given by the axionic scalars 
$a^j$ that stem from reducing the $RR$ $2$--form 
$C_2$ on the $2$--cycles $T^{2,j}$: $a^j=\int_{T^{2,j}} C_2$.
However, some the following formulae take a more suggestive form in terms of 
the geometric moduli $T^j$ as defined above.

The internal antisymmetric tensors $b^j$ and gauge flux $F^j$ are assumed to be 
block--diagonal w.r.t. the three tori, too:
\eqn\internalflux{
F^j\equiv \pmatrix{0& F^j\cr  -F^j& 0}\ \ ,\ \ b^j\equiv \pmatrix{0&b^j\cr-b^j&0}\ .}
Hence, in the $j$--th torus $T^{2,j}$ the total internal anti--symmetric background is
\eqn\totalflux{
\Fc^j=b^j+2\pi\ap F^j=\pmatrix{0& f^j\cr  -f^j& 0}\ ,}
with the value
\eqn\flux{
f^j=b^j+2\pi\ap F^j\ .}

To obtain a gauge theory in four space--time dimensions, we wrap the
$Dp$--brane ($p=9,7,5,3$) on a factorizable $p-3$--cycle $\Cc_{p-3}$, which is the product 
of tori $\Cc_{p-3}=\prod\limits_{j=1}^{(p-3)/2}T^{2,j}$.
Without Wilson lines and additional $U(1)$ gauge fields the  internal dimensions decouple
from the space--time dimensions. Therefore the $(p+1)\times (p+1)$ matrices $G,B$ and
$F$ may be split into the blocks $g_6,b_6,F_6$ and $g,b,F$, describing the internal 
compactification and the $D=4$ space--time, respectively.
The latter shall describe the $D=4$ space--time, with $b=0$.
The matrix $F$ is the anti--symmetric $4\times 4$ matrix
representing the space--time field strength of the gauge field living on the $Dp$--brane.
To this end, after dimensional reduction\foot{The extra factors of $(2\pi)^6$
and $(2\pi)^{p-3}$ stem from the circumference of each circle.} of \Einstein\ on the six--torus 
$T^6=\prod\limits_{j=1}^3 T^{2,j}$ and reducing \BI\ on the $\Cc_{p-3}$--cycle 
specified above we obtain ($p=3,5,7,9$):
\eqn\obtaingauge{\eqalign{
\Sc_{4}&=-\int d^4x\ \sqrt{-g}\ 
\lf\{\fc{e^{-2\phi_{10}}}{(2\pi)\ \ap^4}\ R\ 
\prod_{j=1}^3\sqrt{\det(g^j)}\ri.\cr
&-\lf. T_p\ (2\pi)^{p-3}\ e^{-\phi_{10}}\ 
\sqrt{\det(1+2\pi\ap\ Fg^{-1})}\ \prod_{j=1}^{(p-3)/2}|n^j|\ \sqrt{\det(g^j+\Fc^j)}\ri\}.}}
The integer $n^j$ counts how often the $Dp$--brane is wrapped around the torus
$T^{2,j}$. From 
this action we may read off the gravitational 
coupling\foot{The coefficient of $R$ in $D=4$ is
defined to be $\h\kappa_4^{-2}=(16\pi\ G_N)^{-1}$, with $G_N$ being the $D=4$ Newton constant.} 
$G_N=M^{-2}_{\rm Planck}$ in 
four space--time dimensions ($D=4$)
\eqn\gravitational{
M^2_{\rm Planck}=8\ \fc{e^{-2\phi_{10}}}{\ap^4}\ \prod_{j=1}^3\sqrt{\det(g^j)}=
8\ \fc{e^{-2\phi_{10}}}{\ap^4}\ \prod_{j=1}^3 \im(T^j)=8\ 
\fc{e^{-2\phi_4}}{\ap}\ ,}
\ie $\kappa_4^{-2}=\fc{e^{-2\phi_4}}{\pi\ap}$.
Here we have introduced the dilaton field $\phi_4$ in $D=4$ \MSch 
\eqn\dilfour{
\phi_4=\phi_{10}-\h\ln\lf[\im(T^1)\im(T^2) \im(T^3)/\alpha'^3\ri].}
Furthermore, for the various gauge couplings referring to the $Dp$--brane wrapped on a 
$\Cc_{p-3}$--cycle we get\foot{With $T^j_2=\im(T^j)\ ,\ T^j_1=\re(T^j)$ and 
$U^j_2=\im(U^j)\ ,\ U^j_1=\re(U^j)$.}:
\eqn\gaugecouplings{\eqalign{
g^{-2}_{Dp}&=(2\pi)^{-1}\ \ap^{\fc{3-p}{2}}\ e^{-\phi_{10}}\ 
\prod_{j=1}^{(p-3)/2} |n^j|\ \sqrt{\det(g^j+\Fc^j)}\cr
&=(2\pi)^{-1}\ 
\ap^{\fc{3-p}{2}}\ e^{-\phi_{10}}\ \prod_{j=1}^{(p-3)/2} |n^j|\ |T_2^j+if^j|\cr
&=e^{-\phi_4}\ (2\pi)^{-1}\ \ap^{3-\fc{p}{2}}\ \prod_{j=1}^3\fc{1}{\sqrt{\im(T^j)}}\ 
\prod_{j=1}^{(p-3)/2} |n^j|\ |T_2^j+if^j|\ .}}
As we will see in subsection 3.3. it is the last line of \gaugecouplings, which 
is most directly 
inferred from string amplitudes, \ie disk scattering of two gauge fields and one modulus.
Hence we call the moduli fields $T^j$
and $U^j$ appearing above K\"ahler and complex structure moduli in the {\it
string basis}, respectively.
For any N=1 orientifold/orbifold compactification of \ti with
$D9$-- and $D5$--branes the gauge couplings take the form
\gaugecouplings\ up to some 
normalization constant. Without fluxes according to N=1 SUSY in $D=4$ these couplings have to 
represent real parts of a holomorphic function, namely the gauge kinetic function.
Therefore we introduce new fields $s$ and $t^j$ with their corresponding (real parts) reproducing 
the $D9$ and $D5$--brane couplings without fluxes. These fields are called
the moduli in the {\it supergravity field--theory basis} in contrast to $T^j$ and $U^j$ referring
to the {\it string basis}.
The real part of the dilaton field $s$ gives the coupling of the gauge fields of the $D9$ brane
with wrapping number $n^j=1$ \doubref\ABFPT\AFIV 
\eqn\gaugeDnine{
\re\ s:=g_{D9}^{-2}=(2\pi)^{-1}\ \fc{e^{-\phi_{10}}}{\ap^3}\ T_2^1 T_2^2T_2^3=
(2\pi)^{-1}\ \fc{e^{-\phi_4}}{\ap^{3/2}}\ \sqrt{T_2^1 T_2^2T_2^3}\ .}
The K\"ahler moduli $t^j$ in the {\it field--theory basis} follow from
wrapping once ($n^j=1$) the $D5$ on the tori $T^{2,j}$ \doubref\ABFPT\AFIV:
\eqn\gaugeDfive{
\re\ t^j:=g_{D5,j}^{-2}=(2\pi)^{-1}\ \fc{e^{-\phi_{10}}}{\ap}\ T_2^j=
(2\pi)^{-1}\ \ap^{1/2}\ e^{-\phi_4}\ \sqrt{\fc{T_2^j}{T_2^k T_2^l}}\ \ ,\ \ 
(j,k,l)=\overline{(1,2,3)}\ .}
The imaginary parts $a,a^j$ of the fields $s$ and $t^j$ describe the corresponding axions following
from the $RR$ form couplings $C_6\wedge F\wedge F$ on the $D9$ and 
$C_2\wedge F\wedge F$ on the $D5$--brane after integrating the $RR$ forms
$C_6$ and $C_2$ over the $6$--cycle $T^6$ and $2$--cycle $T^{2,j}$, respectively:
$a^j=\int_{T^{2,j}} C_2,\ a=\int_{T^6} C_6$ (\cf \eqq \stringmoduli).
The K\"ahler potential for the closed string moduli is \ABFPT\ 
\eqn\kpstu{
\kappa_4^2\, \hat K(M,\ov M)=-\ln(s+\ov s)-\sum_{i=1}^3\ln(t^i+\ov t^i)-\sum_{i=1}^3
\ln(u^i+\ov u^i)\ ,}
where $u^i=-iU^i$ are the complex structure moduli.

In the following we shall concentrate on the gauge coupling of $D9$--branes
with non--vanishing fluxes $F^j$:
\eqn\gaugenine{
g_{D9}^{-2}=\fc{ e^{-\phi_4} }{(2\pi)\ \ap^{3/2} }\ 
\prod_{j=1}^{3} \fc{|n^jT^j+m^j|}{\sqrt{\im(T^j)}}\ .}
Recall that in this equation the real part of $T^j$ is given in terms of the NS
background field $b^j$, just like for the geometric K\"ahler modulus
in heterotic compactifications.
Obviously, the gauge coupling does not depend on the complex structure moduli $U^j$,
in accordance with the general arguments of \Doug.
The $D=4$ gauge couplings following for the $D7,D5$ and $D3$--branes (encoded in \gaugecouplings)
may be obtained from this formula by taking the respective limits $f^j=b^j+2\pi\ap F^j\ra
\infty,\ n^j\ra 0$, and multiplying the result by factors of $(2\pi)^2$. This limit 
converts the Neumann boundary conditions of the coordinates of torus $T^{2,j}$ into Dirichlet.
In the case of N=1 supersymmetry in $D=4$ the fluxes $f^j$, specified in \flux, have to obey the
condition \doubref\BDL\IRAN
\eqn\fluxcondition{
\sum_{j=1}^3\fc{f^j}{\im(T^j)}=\prod_{j=1}^3\fc{f^j}{\im(T^j)}\ .}
With this constraint the gauge coupling \gaugenine\ becomes the real part 
of a holomorphic
function as dictated by N=1 supersymmetry \IBA: 
\eqn\susygaugenine{
g_{D9}^{-2}=|n^1 n^2 n^3|\ 
\lf|\re\lf(s -\ap^{-2}f^1f^2\ t^3-\ap^{-2}f^1f^3\ t^2-\ap^{-2}f^2f^3\
t^1\ri)\ri|\ .}

A similar discussion follows for supersymmetric orientifold/orbifold compactifications of \tb\ 
with $D7$-- and $D3$--branes\foot{In this setup we have: $\re\ s:=g_{D3}^{-2}=
\fc{e^{-\phi_4}}{2\pi}\ \fc{\ap^{3/2}}{\sqrt{T_2^1T_2^2T_2^3}}$, and
$\re\ t^j:=g_{D7,j}^{-2}=\fc{e^{-\phi_4}}{2\pi\ap^{1/2}}\ \sqrt\fc{T_2^kT_2^l}{T_2^j}$.
The gauge coupling of a $D7$--brane, wrapped on the two tori $T^{2,k}$ and $T^{2,l}$ with the 
two--form fluxes $f^k$ and $f^l$ becomes: 
$g_{D7,j}^{-2}=|n^k n^l|\ |\re(t^j-\ap^{-2}\ f^k f^l\ s)|$. The N=1 supersymmetry condition
demands: $\fc{f^k}{T_2^k}=-\fc{f^l}{T_2^l}$. The imaginary parts $a,a^j$ of 
the fields $s$ and $t^j$ describe the corresponding axions following
from the $RR$--form couplings $C_0\wedge F\wedge F$ on the $D3$ and 
$C_4\wedge F\wedge F$ on the $D7$--brane after integrating the $RR$--form
$C_4$ over the $4$--cycle $T^{2,k}\times T^{2,l}$: 
$a^j=\int\limits_{T^{2,k}\times T^{2,l}} C_4$. }.

\subsec{Gauge couplings of rotated $D6$--branes (Type $IIA$ picture)}

Now we perform a $T$--duality in the $y$--directions of each torus $T^{2,j}$.
It implies that the type $IIB$ moduli $T^j,U^j$ are related to the type $IIA$
moduli ${T'}^j,{U'}^j$ as follows:
\eqn\duality{
T^j= -\fc{\ap}{{U'}^j}\ \ \ ,\ \ \ U^j= -\fc{\ap}{{T'}^j}\ \ \ ,\ \ \ j=1,2,3\ .}
Under this duality the $D=10$ coupling constant transforms
according to:
\eqn\transtendil{
e^{-\phi_{10}}\lra \fc{e^{-\phi_{10}}}{\ap^{3/2}}\ 
\prod_{j=1}^3  |{U'}^j|\  \sqrt{\fc{{T'_2}^j}{{U'_2}^j}}\ .}
Note, that the $D=4$ dilaton $\phi_4$ stays  inert under these transformations.
The $D9$--brane with the background gauge fluxes $\Fc^j$ is converted into a $D6$--brane
at angles $\th^j$ w.r.t.  the $x$--axis. The angle $\th^j$ is given by \IRAN
\eqn\anglerelb{
\tan(\pi\th^j)=\fc{f^j}{\im(T^j)}=\fc{1}{\im(T^j)}\ (b^j+2\pi\ap F^j)\ ,}
expressed in \tb quantities. 
After the $T$--duality \duality\ the choice
\eqn\fluxchoice{
F^j=\fc{1}{2\pi\al'}\ \fc{m^j}{n^j}} 
leads to the corresponding \ta relation:
\eqn\anglerel{
\tan(\pi\th^j)=-\fc{{U'}_1^j}{{U'_2}^j}+\fc{m^j}{n^j}\ \fc{|{U'}^j|^2}{{U'}_2^j}=
\fc{\tilde m^j}{n^j}\ \fc{|{U'}^j|^2}{{U'}_2^j}\ .}
Here we introduced the shifted wrapping number
\eqn\shifttilted{
\tilde m^j=m^j-\fc{{U'_1}^j}{|{U'}^j|^2}\ n^j}
describing a tilted torus.
The supersymmetry condition \fluxcondition\ on the fluxes translates into a relation
between the complex structure moduli
\eqn\susysixcomplex{
\sum_{j=1}^3\fc{\tilde m^j}{n^j}\ \fc{|{U'}^j|^2}{{U'_2}^j}=
\prod_{j=1}^3\fc{\tilde m^j}{n^j}\ \fc{|{U'}^j|^2}{{U'_2}^j}\ ,}
which after \anglerel\ becomes a condition on the angles $\th^j$ \BDL:
\eqn\susy{
\th^1+\th^2+\th^3=0\ \ \mod\ \ 2\ .}
Apparently, the duality action \duality\ converts \gaugenine\ into:
\eqn\gaugesix{
g_{D6}^{-2}=\fc{e^{-\phi_4}}{(2\pi)}\ 
\ \prod_{j=1}^{3} \fc{|n^j-m^j{U'}^j|}{\sqrt{\im({U'}^j)}}\ .}
Evidently, the gauge coupling,
being proportial to the volume of
the relevant 3-cycle, does not depend on the $IIA$ K\"ahler moduli ${T'}^j$.
This expression may be directly derived from dimensional reducing the Born--Infeld action \BI\ 
of a $D6$--brane on a $3$--cycle, which is a direct product of three $1$--cycles 
$\Cc^j_1\ ,\ j=1,2,3$.
Each of this $1$--cycle has the wrapping numbers $(n^j,m^j)$ w.r.t. the homology basis of the 
torus $T^{2,j}$.

The real parts of the \tb field theoretical moduli fields $s$ and $t^j$ become after the
$T$--duality transformations \duality\ and \transtendil
\eqn\fieldtypea{
\re\ s'=(2\pi)^{-1}\ e^{-\phi_4}\ \fc{\sqrt{{U'_2}^1{U'_2}^2{U'_2}^3}}{|{U'}^1{U'}^2{U'}^3|}
\ \ \ ,\ \ \ 
\re\ {u'}^j=(2\pi)^{-1}\ e^{-\phi_4}\ \sqrt{\fc{{U'_2}^j}{{U'_2}^k{U'_2}^l}}\ 
\lf|\fc{{U'}^k{U'}^l}{{U'}^j}\ri|\ ,}
which respectively boil down to $\re\ s'=\fc{e^{-\phi_4}}{(2\pi)}\ 
\fc{1}{\sqrt{{U'_2}^1{U'_2}^2{U'_2}^3}}$
and $\re\ {u'}^j=\fc{e^{-\phi_4}}{(2\pi)}\ \ \sqrt{\fc{{U'_2}^k{U'_2}^l}{{U'_2}^j}}$ 
for rectangular tori ${\al'}^j=\pi/2$.
With the additional constraint \susysixcomplex\ the gauge coupling \gaugesix\ becomes
the real part of a holomorphic function:
\eqn\susygaugesix{
g_{D6}^{-2}=
\lf|\re\lf(n^1n^2n^3\ s'-n^1\tilde m^2\tilde m^3\ {u'}^1-
n^2\tilde m^1\tilde m^3\ {u'}^2-n^3\tilde m^1\tilde m^2\ {u'}^3\ri)\ri|\ .}

\subsec{Scattering of two gauge fields and one modulus from $D$--branes with fluxes}

In this section we want to check the formulae \gaugenine\ and \gaugesix\ by performing
an explicit string calculation. Let us start on the \tb or \ti side with a 
stack of $D9$--branes describing the generic gauge group $G_a$. We allow for non--vanishing
fluxes $\Fc^j=b^j+2\pi\ap F^j$ w.r.t. the internal torus dimensions on which the $D9$ is wrapped.
As already mentioned before, we may obtain lower dimensional branes by taking the special 
limit $\Fc^j\ra\infty$ in some planes $j$.
To gain the full moduli dependence of the tree--level gauge couplings, we calculate
the disk amplitudes $\vev{A^a A^a V_{T^j}}$ and $\vev{A^a A^a V_{U^j}}$
involving two gauge fields $A^a_\mu$ 
on the boundary and one closed string modulus $T^j$ and $U^j$, respectively.
The latter refer to the $j$--th subtorus $T^{2,j}$.
In the zero ghost picture the gauge field vertex operator is
\eqn\gaugevertex{
V^{(0)}_{A^a}(z,k)=\lambda^a\ \xi_\mu\ \lf[\p X^\mu+i\ (k\psi)\ \psi^\mu\ri]\ 
e^{ik_\rho X^\rho(z)}\ .}
Here $\lambda^a$ is a Chan--Paton factor in the adjoint of the gauge group $G_a$,
describing the fundamental gauge degrees of freedom at the open string ends. 
In addition, the polarization $\xi_\mu$ has to fulfill $\xi_\mu k^\mu=0$.
Furthermore the $T^j$ and $U^j$ vertex operators in the $(-1,-1)$ ghost picture are given by
\eqn\vertexTU{\eqalign{
V_{T^j}^{(-1,-1)}(\ov z,z,k)&=\h\ e^{-\tilde\phi(\ov z)}\ e^{-\phi(z)}\ 
\fc{\p}{\p T^j}\ (g^j+b^j)_{kl}\ \tilde\psi^k(\ov z)\ \psi^l(z)\
e^{ik_\rho X^\rho(z,\ov z)}\cr
&=\fc{1}{T^j-\ov T^j}\ e^{-\tilde\phi(\ov z)}\ e^{-\phi(z)}\ \tilde\Psi^j(\ov z)\ 
\ov{\Psi^j}(z)\ e^{ik_\rho X^\rho(z,\ov z)}\ ,\cr
V_{U^j}^{(-1,-1)}(\ov z,z,k)&=\h\ e^{-\tilde\phi(\ov z)}\ e^{-\phi(z)}\ 
\fc{\p}{\p U^j}\ (g^j+b^j)_{kl}\ \tilde\psi^k(\ov z)\ \psi^l(z)\
e^{ik_\rho X^\rho(z,\ov z)}\ ,\cr
&=\fc{-1}{U^j-\ov U^j}\ e^{-\tilde\phi(\ov z)}\ e^{-\phi(z)}\ 
\tilde\Psi^j(\ov z)\ \Psi^j(z)\ e^{ik_\rho X^\rho(z,\ov z)}\ ,}}
with the backgrounds\foot{Note, that for simplicity we assume a  compactification 
on a six--torus $T^6$ with the latter being a direct product of three single two--tori $T^{2,j}$.}
$g^j$ and $b^j$ defined in \metrics. More precisely the vertex operator for
the imaginary part $T_2^j$ of $T$ 
is given by $V_{T_2}=i(V_T-V_{\ov T})$, which
amounts to symmetrizing the vertex operator $V_T$ w.r.t. to the left- and 
right-movers. The RR vertex operator for the real part of $T$ can be obtained
from space-time supersymmetry.
In the following we compute the amplitudes for the imaginary part 
$T_2$ by using the above operator $V_T$ and its conjugate 
and summing the two amplitudes
at the end of the computation.

Above we have also introduced the complex bosonic and fermionic fields
($j=1,2,3$):
\eqn\complexify{\eqalign{
\ov Z^j&=\sqrt{\fc{T^j_2}{2 U^j_2}}\   (X^{2j-1}+U^jX^{2j})\ \ \ ,\ \ \ 
Z^j=\sqrt{\fc{T^j_2}{2 U^j_2}}\ \      (X^{2j-1}+\ov U^j X^{2j})\ ,\cr
\ov\Psi^j&=\sqrt{\fc{T^j_2}{2 U^j_2}}\ (\psi^{2j-1}+U^j\psi^{2j})\ \ \ ,\ \ \ 
\Psi^j=\sqrt{\fc{T^j_2}{2 U^j_2}}\     (\psi^{2j-1}+\ov U^j\psi^{2j})\ .}}
In this writing, the  Green's functions \green\ for the internal bosonic fields
$\p Z$ and fermions $\Psi$ take the simple form:
\eqn\simplegreen{\eqalign{
&\vev{\p Z^j(z_1)\ \p \ov Z^j(z_2)}=-\fc{1}{(z_1-z_2)^2}\ \ \ ,\ \ \ 
\vev{\p Z^j(z_1)\ \p Z^j(z_2)}=0\cr
&\vev{\Psi^j(z_1)\ \ov \Psi^j(z_2)}=\fc{1}{z_1-z_2}\ \ \ ,\ \ \ 
\vev{\Psi^j(z_1)\ \Psi^j(z_2)}=0\ .}}
Moreover, the Green's functions \greeni\ assume a very compact form, too.
First, the $D$--matrix, introduced in \Dmatrix, 
takes for the metric $g^j$ and the antisymmetric background $\Fc^j=b^j+2\pi\ap F^j$ 
(\cf \totalflux) the form\foot{Note, that in two uncompactified directions, the matrix $D$
takes the familiar form: $D^{\mu\nu}=\fc{1}{1+f^2}\ \pmatrix{1-f^2&-2f\cr
                                                         2f&1-f^2}$, which is orthogonal $DD^t=1$.
However, we take $D^{\mu\nu}=\delta^{\mu\nu}$ in the four uncompactified directions.}:
\eqn\internalD{\eqalign{
D^{j,ab}&=\fc{2}{(T^j-\ov T^j)(U^j-\ov U^j)}\ \fc{1}{(T^j-\ov T^j)^2-4\ (f^j)^2}\times\cr 
&\hskip-1.2cm
\pmatrix{-2\ |U^j|^2\ [4\ (f^j)^2+(T^j-\ov T^j)^2]& 
U(2f^j+T^j-\ov T^j)^2+\ov U^j(2f^j-T^j+\ov T^j)^2\cr
U(2f^j-T^j+\ov T^j)^2+\ov U^j(2f^j+T^j-\ov T^j)^2&
-2\ [4\ (f^j)^2+(T^j-\ov T^j)^2]}.}}
With this matrix, we obtain\foot{Note: $\vev{\p \ov Z^j(z_1)\ \ov\p Z^j(\ov z_2)}=
-\fc{\ov{D^j}}{(z_1-\ov z_2)^2}$,\ $\vev{\ov\p Z^j(\ov z_1)\ \p \ov Z^j(z_2)}=
-\fc{\ov{D^j}}{(\ov z_1-z_2)^2}$, and:\br 
$\vev{\ov \Psi^j(z_1)\ \tilde\Psi^j(\ov z_2)}=\fc{\ov D^j}{z_1-\ov z_2}$\ ,
$\vev{\tilde\Psi^j(\ov z_1)\ \ov \Psi^j(z_2)}=\fc{\ov D^j}{\ov z_1-z_2}$\ ,
$\vev{\ov{\tilde\Psi}^j(\ov z_1)\ \Psi^j(z_2)}=\fc{D^j}{\ov z_1-z_2}$\ .} 
for \greeni:
\eqn\simplegreeni{\eqalign{
&\vev{\p Z^j(z_1)\ \ov\p \ov Z^j(\ov z_2)}=-\fc{D^j}{(z_1-\ov z_2)^2}\ \ \ ,\ \ \ 
\vev{\p Z^j(z_1)\ \ov\p Z^j(\ov z_2)}=0\ ,\cr
&\vev{\Psi^j(z_1)\ \ov{\tilde \Psi}^j(\ov z_2)}=\fc{D^j}{z_1-\ov z_2}\ \ \ ,\ \ \ 
\vev{\Psi^j(z_1)\ \tilde \Psi^j(\ov z_2)}=0\ ,}}
with: 
\eqn\complexD{
D^j=\fc{T^j-\ov T^j+2\ f^j}{T^j-\ov T^j-2\ f^j}=\fc{T_2^j-i\ f^j}{T_2^j+i\ f^j}\ .}
Obviously, we have $D^j \ov D^j=1$, and
\eqn\DN{
D^j=\cases{1\ , & {\rm pure Neumann\ ,}\cr
          -1\ , & {\rm pure Dirichlet\ ,}}}
for boundary conditions of the same sort in both directions of the $T^{2,j}$.
These two cases are  simply obtained from
\complexD\ in the limit $f^j\ra 0$ and $f^j\ra\infty$, respectively.
Due to internal $U(1)$ charge conservation, correlators involving fields from different 
planes $j$ vanish for our choice of background,
the six--torus $T^6$ being a direct product of three single two--tori $T^{2,j}$.

First we shall calculate the amplitude: 
\eqn\todoa{
\Ac_{A^{a_1} A^{a_2} T^j}=\fc{i}{4\pi}\, \int \fc{dz_1 dz_2 d^2z_3}{V_{\rm CKG}}\ 
\vev{V^{(0)}_{A^{a_1}}(z_1,k_1)\ V^{(0)}_{A^{a_2}}(z_2,k_2)\ 
V_{T^j}^{(-1,-1)}(\ov z_3,z_3,q)}\ .}
We have chosen the modulus vertex operator \vertexTU\
in the $(-1,-1)$ ghost picture in order to guarantee a total ghost charge of $-2$ on the disk.
This requirement is a consequence of the superdiffeomorphism invariance on the string world sheet.
We extract from \todoa\ the kinematics 
$\kappa=\h[(p_1p_2)(\xi_1\xi_2)-(p_1\xi_2)(p_2\xi_1)]$ following from
the gauge kinetic term $\fc{1}{4}F^{a_1}_{\mu\nu}F^{a_2,\mu\nu}$ in four space--time dimensions.
Here $V_{\rm CKG}$ is the volume of the conformal Killing group $PSL(2,\IR)$, which
leaves the boundary ($\im(z)=0$) of the disk fixed.
The correlators appearing in \todoa\ are of the form described in section 2. 
According to \eqq \closedm, the closed string momentum $q$ is split into 
$q^{||}=\h(q+Dq)=\h(k_3+k_4)$.
The contraction of the exponentials in \todoa\ yields
\eqn\expo{\eqalign{
\Ec:=\vev{e^{ik_1X(z_1)}\ e^{ik_2X(z_2)}\ e^{ik_3X(\ov z_3)}\ e^{ik_4X(z_3)}}&=
|z_1-z_2|^{s}\ |z_1-\ov z_3|^t\ |z_1-z_3|^u\cr
&\times|z_2-\ov z_3|^u\ |z_2-z_3|^t\ |\ov z_3-z_3|^s\ .}}
The last expression is subject to the momentum constraints \momentum and \cons.
Due to $PSL(2,\IR)$ invariance on the disk, we may fix three vertex positions. A convenient choice
for the kind of setup we are considering (\cf \HKi) is\foot{With this choice
the correlator \expo\ becomes \eqn\expob{\Ec=\lf(\fc{x^2+1}{4x}\ri)^{2t}\ .}}:
\eqn\position{
z_1=x\ \ ,\ \ z_2=-x\ \ ,\ \ z_3=i\ \ ,\ \ \ov z_3=-i\ .}
This choice implies the $c$--ghost contribution:
\eqn\cghost{
\vev{c(z_2)\ c(z_3)\ \tilde c(\ov z_3)}=(z_2-z_3)(z_2-\ov z_3)(z_3-\ov z_3)=2i\ (1+x^2)\ .}
After including this correlator, we obtain:
\eqn\findgauge{\eqalign{
\Ac_{A^{a_1} A^{a_2} T^j}&=\fc{i}{4\pi}\ 2^{-1-4t}
\ \fc{\ov D^j}{T^j-\ov T^j}\lf\{(\xi_1\xi_2)\ (1+2t)\ 
\fc{i}{4}\  
\int_{0}^\infty dx\ \fc{(x^2+1)^{2t+1}}{x^{2t+2}}\ri. \cr
&+\lf[(k_3\xi_1)(k_4\xi_2)
\int_{0}^\infty dx\ \fc{(x+i)^{2t+1}(x-i)^{2t-1}}{x^{2t+1}}\ri.\cr
&\lf.\lf.-(k_4\xi_1)(k_3\xi_2)
\int_{0}^\infty dx\ \fc{(x+i)^{2t-1}(x-i)^{2t+1}}{x^{2t+1}}\ri]\ri\}\ .}}
The relevant integrals may be performed with the more general formula
\eqn\integralii{\eqalign{
I(\delta,\alpha)&=\int\limits_0^\infty dx\ x^{\delta-1}\ (x-i)^{\alpha-\delta}\ 
(x+i)^{-\alpha-\delta}\cr
&=\sqrt\pi\ 2^{-\delta}\ e^{-\h\pi i\alpha}\ 
\fc{\Gamma\lf(\fc{\delta}{2}\ri)\ \Gamma\lf(\h+\fc{\delta}{2}\ri)}
{\Gamma\lf(\h+\fc{\delta}{2}-\fc{\al}{2}\ri) \Gamma\lf(\h+\fc{\delta}{2}+\fc{\al}{2}\ri)} ,}}
which will be relevant in the following.
To this end, we arrive at:
\eqn\finalgauge{\eqalign{
\Ac_{A^{a_1} A^{a_2} T^j}&=\fc{1}{4}\ 
\fc{\ov D^j}{T^j-\ov T^j}\ \lf\{\ 
t\ (\xi_1\xi_2)+\lf[\ (k_3\xi_1)(k_4\xi_2)+(k_4\xi_1)(k_3\xi_2)\ \ri]\ \ri\}\ 
t\ \fc{\Gamma(-2t)}{\Gamma(1-t)^2}\cr
&=\fc{i\ \ov D^j}{2\ T_2^j}\ \lf[\ \h(p_1p_2)(\xi_1\xi_2)-\h(p_1\xi_2)(p_2\xi_1)\ \ri]\ 
t\ \fc{\Gamma(-2t)}{\Gamma(1-t)^2} \ .}}
Due to 
$t\ \fc{\Gamma(-2t)}{\Gamma(1-t)^2}=-\h+\Oc(k^2)$, at second order in the momenta
we obtain from \finalgauge\  
the gauge kinematics $\kappa$ of two gauge bosons.

To compare with the field theory, the gauge field vertices have to be rescaled
as $V_{A^a}\ra g_a^{-1}\ V_{A^a}$. 
This manipulation makes sure, that the string amplitude \todoa,  
calculated with the gauge vertices in the string basis \gaugevertex,  
describes a field theoretical quantity. The leading term of the 
L/R symmetrized amplitude can then be equated with the $T^j_2$--derivative
\eqn\dgli{
\Ac_{A^{a_1} A^{a_2} T_2^j}=
i(\Ac_{A^{a_1} A^{a_2} T^j}-\Ac_{A^{a_1} A^{a_2} \ov T^j})=
\kappa\, \fc{1}{4}\fc{D^j+\ov D^j}{T^j_2}g_a^{-2}=
\kappa\, \p_{T_2^j}\ g^{-2}_a.}
On the other hand, the amplitudes  
$\Ac_{A^{a_1} A^{a_2} U^j}$ and $\Ac_{A^{a_1} A^{a_2} \ov U^j}$
give a vanishing result due to internal charge conservation. A result essentially following
from the correlator $\vev{\Psi^j(z_1)\ \tilde \Psi^j(\ov z_2)}=0$ in \simplegreeni.
This yields the additional differential equations:
\eqn\dgliii{
\p_{U^j}\ g^{-2}_a=0\ \ \ ,\ \ \ \p_{\ov U^j}\ g^{-2}_a=0\ \ \ ,\ \ \ j=1,2,3\ .}
The equations \dgli\ and \dgliii\ have  the solution:
\eqn\solutiongauge{
g^{-2}_a={\rm const.}\ \ e^{-\phi_4}\ \prod_{j=1}^3
\fc{\sqrt{(T^j_2)^2+(f^j)^2}}{\sqrt{T^j_2}}.}
We included the dilaton factor arising in the path integral of any string scattering
on the disk. The latter has Euler number $\chi=1$. 
This dilaton dependence
may be explicitly checked by calculating a scattering of two gauge fields and the dilaton field
$\phi_4$ on the disk (\cf Appendix \appB).

Alternatively working instead with the moduli \vertexTU\ in the real basis we obtain the 
differential equation:
\eqn\realgauge{\eqalign{
\p_{T_2^j}\ g^{-2}_a&=\fc{1}{4}\ \Tr\lf\{\fc{\p}{\p T_2^j} (g^j+b^j)\ D^j\ri\}\ g^{-2}_a\cr
&=\fc{1}{4}\ \Tr \lf\{\fc{\p}{\p T_2^j} (g^j+b^j)\ [-(g^j)^{-1}+2\ (g^j+\Fc^j)^{-1}]\ri\}\ g^{-2}_a
\ \ \ ,\ \ \ j=1,2,3\ ,}}
which may be cast into:
\eqn\realgaugei{
\p_{T_2^j}\ \ln g^{-2}_a=\h\ \p_{T_2^j}\ \ln\lf\{\ \det\lf[1+(b^j+2\pi\ap\ F^j) (g^j)^{-1}\ri]\ 
\sqrt{\det g^j}\ \ri\}\ .}
from which one concludes:
\eqn\concludegauge{
 g^{-2}_a={\rm const.}\ e^{-\phi_4}\  (\det\ g)^{-1/4}\ \sqrt{\det(g+b+2\pi\ap F)}\ ,}
in precise agreement with \gaugecouplings\ for $p=9$.
It is quite remarkable, that our disk calculation \todoa, which involves both 
open and closed string states and which borrows quantities \Dmatrix\ 
from non--commutative field theory, precisely reproduces the coupling \gaugenine\
following from the Born--Infeld action \BI.

\subsec{Scattering of two gauge fields and one modulus from $D$--branes at angles}

Our findings \dgli\ and \dgliii, describing the moduli dependence of gauge
couplings of $D9$--branes with fluxes $F^j$, directly translate into the 
the $T$--dual picture \duality\ of $D6$--branes at angles $\th^j$ (\cf \fluxchoice):
\eqn\Dgli{\eqalign{
\p_{{U'_2}^j}\ g^{-2}_a&=-\fc{1}{4U^{'j}_2}\ \lf(
\fc{n^j-m^j\ {\ov U'}^j}{n^j-m^j\ {U'}^j}+{\rm c.c.}\ri)\ \ g^{-2}_a\ ,\cr
\p_{{T'}^j}\ g^{-2}_a&=0\ \ \ ,\ \ \ \p_{\ov {T'}^j}\ g^{-2}_a=0\ \ \ ,\ \ \ j=1,2,3\ .}}
These differential equations just follow from \dgli\ and \dgliii\ after performing
the duality transformation \duality. They integrate to \gaugesix.

On the other hand, an explicit calculation of the string disk amplitudes 
$\vev{A^a A^a U_2'}$ and $\vev{A^a A^a T_2'}$ involving the dual moduli fields
$\Tp$ and $\Up$ should also yield the above equations. 
In the following we give a derivation of the
CFT correlators for branes at angles. It will become obvious that 
the $IIA$ computation indeed yields \Dgli. In addition the discussion 
leads to a simple interpretation 
of the $D$ matrices in the $f\neq 0$ correlators in \simplegreeni,
that have been obtained above by transforming the correlators of \doubref\ACNY\SW\ 
to the complex basis.

The moduli on a single torus $T^2$ with radii $R_i$ are defined
in \stringmoduli. In the following we drop the primes on the $IIA$ moduli for
simplicity. Let $Y^a$ denote the two bosons corresponding to the 
lattice basis, $Y^a \sim Y^a+1$. They have the correlation functions
$$
\langle Y^a(z) Y^b(w) \rangle =  -g^{ab}\ln|z-w|^2,
\qquad g^{ab}=\fc{1}{T_2U_2}\pmatrix{|U|^2&-U_1\cr -U_1&1}.
$$
The complex coordinates on $T$ defined in \complexify
$$
Z=\sqrt{\fc{T_2}{2U_2}}(Y^1+\ov U Y^2),
$$
have correlators $\langle Z(z) \ov Z(w) \rangle = -\ln|z-w|^2$. The coordinates $Z$ are related
to any orthogonal basis $X^a$ by 
\eqn\ortji{
Z= \fc{\delta}{\sqrt{2}}(X^1+iX^2),
}
where $\delta=e^{i\al}$ is a yet arbitrary phase factor.

On the double cover of the disc, the interactions between 
left- and right-movers at the boundary are given by 
$\langle \p X^a(z)\ov\p X^a(\ov w)\rangle = D^a \langle \p X^a(z)\p X^a(\ov w)\rangle$, 
where $D^a=\pm 1$ in the $N$ and $D$ directions, respectively. 
We may fix the coordinates such that $X^2$ is the $D$ direction, so 
the $D1$--string wraps along $X^1$. The non-vanishing correlators are then:
\eqn\cfi{
\langle  \p Z(z)\,  \p \ov Z(w)\rangle   = -\fc{1}{(z-w)^2},\qquad
\langle  \p Z(z)\, \ov \p Z(\ov w) \rangle  = -\fc{\delta^2}{(z-\ov w)^2},}
and their complex conjugates. Note the non-vanishing correlator of
two holomorphic coordinates on the torus, induced by the Dirichlet
boundary.

We see that in the complex basis, the effect of the mixed boundary conditions
on the CFT computation is quite trivial when using holomorphic fields $\p Z$ on the double. 
The computation essentially reduces to the pure Neumann case after the replacement
$\ov \p \ov Z(\ov w) \to \p Z(\ov w)$, up to a simple product of
phase factors $\delta$, collected from  the vertex operators.
By supersymmetry, a similar comment applies to the correlators of the
fermionic superpartners $\Psi$. 

The angle of the $D1$--brane in the lattice basis 
can be identified by rewriting the $X^a$ in terms of the 
$Y^a$. A string  wrapping the class $n e_1 +m e_2$ of the lattice
corresponds to the ratio $F=m/n$ of the coefficients
of $Y^2$ and $Y^1$ in $X^1$, respectively.
In this way one recovers the definitions 
$\tan(\al)=\fc{f}{U_2}$ with $f=F-U_1.$
Since the above describes the correlators for a $D1$--brane wrapped on $T^2$, 
or, in the dual language, correlators in the presence of non-zero
$B_{NS}$ fields, \eqq \cfi\ must be simply the non-commutative 
correlators of \doubref\ACNY\SW, written in the natural complex basis
on the torus. This fact is verified by \simplegreeni, using $\ov
D=\delta^2$
and noting that the $T$--duality transformation $X^2_R \to -X^2_R$ implies the
replacement $\ov \p Z$ by $\ov \p \ov Z$ when going from $IIA$ to $IIB$. 
With these replacements, the $IIB$ vertex operators $V_{T/U}$ and their amplitudes
immediately translate to the $T$--dual $IIA$ vertex operators $V_{U'/T'}$ and their 
amplitudes, and eventually lead to the $T$--dual equations \Dgli. In the
following we will sometimes 
use this simple direct relation between $IIA$ and $IIB$ 
amplitudes to switch between the two $T$-dual descriptions.

\newsec{Untwisted matter metric}

In this section, we consider tree--level disk scattering amplitudes involving
matter fields inserted at the boundary of the disk and the bulk 
moduli fields.
We shall determine  the metric $G_{C_i\ov C_k}$ for untwisted charged 
matter fields $C_i$ and $\ov C_k$.
In order to obtain a chiral N=1 theory in $D=4$ \tb or \ti, 
the $T^6$--torus is modded out by some orbifold group. 
The details of the latter are not relevant for untwisted matter fields originating
from the untwisted sector.
Generically in \tb or \ti there appear untwisted charged matter fields  $C^9_i,\ i=1,2,3$ 
from open strings starting and ending on $D9$--branes. 
They are charged under the gauge group resulting from the $D9$--branes.
Similarly, open strings starting and ending on the same $D5_j$--brane give rise to the 
matter fields $C^{5_j}_i,\ i,j=1,2,3$. 
Here the index $j$ denotes the torus $T^{2,j}$, on which the $D5$ is
wrapped. The open strings are only charged w.r.t. the gauge group from the $D5_j$.

Before we start let us add the following comment:
in the heterotic string the corresponding 3--point amplitudes on the sphere
between two matter fields and one modulus vanish due to internal charge
conservation.
Therefore in the heterotic string one has to compute 4--point amplitudes 
between two matter fields and
two moduli in order to extract non-trivial informations about the heterotic
matter field K\"ahler metrics. 
This situation now changes in type $II$ models with closed and open string
fields.
Specifically, as we will now see, some 3--point amplitudes between two open
string matter fields and one closed string modulus are allowed by internal
charge conservation. This essentially stems from the fact that the closed
string vertex operator can be treated as two open string vertex operators,
\ie one is basically computing a 4--point amplitude. 
However one has to be quite careful, since there can be in general additional
moduli dependences for the matter field K\"ahler metrics that are not captured
by the 3--point amplitudes.
To obtain these further informations we will also calculate 4--point
amplitudes with two moduli and two matter fields.

\subsec{Three--point amplitudes}

As in the case of the gauge couplings we first work on the \tb side,
where we compute the scattering 
of two matter fields $C_i, \ov C_k$ and the closed string modulus $T^j$.
The open string $\sigma$--model action has the boundary term $\int dz\ A_i\ \p X^i$, which
may be written in complex coordinates \complexify\ as
\eqn\boundary{
\sum_{i=1}^6 A_i\ \p X^i=
\sum_{j=1}^3 \fc{\sqrt 2}{(U^j-\ov U^j)^{1/2}(T^j-\ov T^j)^{1/2}} 
\lf[C_j\ \p Z^j-\ov C_j\ \p\ov Z^j\ri]\ ,}
with $C_i=A_{2i-1}+U^i\ A_{2i}\ ,\ i=1,2,3$.
From this term the vertex operator for untwisted matter fields $C_i$ in the zero ghost picture
may be read off\foot{We discard the normalization $\fc{\sqrt 2}{(U-\ov U)^{1/2}(T-\ov T)^{1/2}}$,
in order for the string vertex to yield canonical normalized kinetic energy in the string basis.}:
\eqn\zeromatter{
V_{C_i}^{(0)}(z,k)=\lambda\ \lf[\p Z^i+i(k\psi)\Psi^i\ri]\ e^{ik_\nu X^\nu(z)}\ .}
The latter is inserted at the boundary of the disk. 
This vertex operator involves untwisted fields from the internal torus directions inserted at
the boundary. 

We must not consider couplings of matter fields $C_i$ and $\ov C_k$ referring to different 
internal complex planes $i$ and $k$. 
Due to internal charge conservation those couplings must vanish 
at second order in the matter fields:
\eqn\conclude{
G_{C_i\ov C_k}=0\ \ \ ,\ \ \ i\neq k\ .}
This can be easily anticipated from the form of the vertex operator \zeromatter.
Hence we shall consider the following three--point amplitude\foot{Compared to \todoa\ we
disregarded a factor of $\h$, which takes into account the factor $\sqrt 2$ in \boundary.} 
\eqn\todoA{
\Ac_{C_i\ov C_iT^j}=\fc{i}{2\pi}\, \int \fc{dz_1 dz_2 d^2z_3}{V_{\rm CKG}}\ 
\vev{V_{C_i}^{(0)}(z_1,k_1)\ V_{\ov C_i}^{(0)}(z_2,k_2)\ V_{T^j}^{(-1,-1)}(\ov z_3,z_3,q)}\ ,}
with the closed string vertex operator \vertexTU\ in the $(-1,-1)$ ghost picture. 
The derivative w.r.t. the physical scalar Im$(T)$ is given by the sum of
\todoA\ and the same amplitude with $T$ replaced by $\ov T$.

There are two contributions to the correlator \todoA: one, denoted by $X_1$,
from contracting the first term of both matter field vertices \zeromatter\ 
and the other, $X_2$, from contracting
the second terms of both matter vertices with $V_{T^j}$.
The correlators to be done are basic and are given in \green\ and \greeni.
To this end we obtain
\eqn\tothisend{\eqalign{
X_1&=\fc{\ov D^j}{T^j-\ov T^j}
\int \fc{dz_1 dz_2 d^2z_3}{V_{\rm CKG}} \Ec\ \fc{-1}{(\ov z_3-z_3)^2\ (z_1-z_2)^2}\ ,\cr
X_2&=\fc{\ov D^j}{T^j-\ov T^j}
\int \fc{dz_1 dz_2 d^2z_3}{V_{\rm CKG}} \Ec\fc{1}{\ov z_3-z_3}\fc{k_1k_2}{z_1-z_2} 
\lf(\fc{1}{(z_1-z_2)(\ov z_3-z_3)}+\fc{\delta^{ij}}{(z_1-z_3)(z_2-\ov z_3)}\ri)}}
with the requirement $s=-2t$.
As in subsection 2.3, we fix three vertex positions according to \position\ and introduce
the correlator \cghost\ to arrive at
\eqn\tothisendi{\eqalign{
X_1&=-\fc{1}{\pi}\ 2^{-4-4t}\ I(-1-2t,0)\ \fc{\ov D^j}{T^j-\ov T^j}=
-\fc{1}{\sqrt\pi}\ 2^{-2t-3}\ \fc{\Gamma(-\h-t)}{\Gamma(-t)}\ \fc{\ov D^j}{T^j-\ov T^j} ,\cr
X_2&=\fc{i}{\pi}\ 2^{-1-4t}\ t\ \lf[\fc{i}{4}\  
I(-1-2t,0)-\delta^{ij}\ I(-2t,1)\ri]\ \fc{\ov D^j}{T^j-\ov T^j}\cr
&=-\fc{2^{-2-2t}}{\sqrt\pi}\  \lf[t-\delta^{ij}(1+2t)\ri]\ \fc{\Gamma(-\h-t)}{\Gamma(-t)}\ 
\fc{\ov D^j}{T^j-\ov T^j}\ ,}}
with the integral $I(\delta,\alpha)$ introduced in \integralii.
Hence the final result for the amplitude \todoA\ is:
\eqn\totalamp{
\Ac_{C_i\ov C_i T^j}=\fc{it}{\sqrt \pi}\ (1-2\delta^{ij})\ \fc{\ov D^j}{T_2^j}
\ 2^{-3-2t}\ \fc{\Gamma(\h-t)}{\Gamma(1-t)}\ .}
For low momenta $t=-2p_1p_2$, 
the result agrees with the $D$--brane effective action for the scalar
matter fields coupling to the closed string modulus $T^j$. 
Therefore, the first term of the expansion 
\eqn\totalami{
\Ac_{C_i\ov C_i T^j}=-i\ (p_1p_2)\ \fc{\ov D^j}{4T_2^j}\ \ 
(1-2\delta^{ij})  +\Oc(t^2)}
describes the 
derivative of the metric $G_{C_i\ov C_i}$ for the scalar matter fields 
$C_i$ and $\ov C_i$. 
Up to a minus sign we obtain the same result for amplitude 
with an insertion of the conjugate vertex operator for $\ov T^j$,
leading to the differential equation
\eqn\dgl{
\p_{T_2^j}\ G_{C_i\ov C_i}=i(\Ac_{C_i\ov C_i T^j}-\Ac_{C_i\ov C_i \ov T^j})
=\fc{D^j+\ov D^j}{4 T^j_2}\, (1-2\delta^{ij})\ G_{C_i\ov C_i}.}
We have changed
the normalization of the matter field vertex operator \zeromatter.
The latter refers to the string basis and leads to canonically normalized matter 
metrics $G_{C_i\ov C_i}\sim 1$. 
In order that our string amplitude reproduces a 
field--theory result, we have to normalize it properly. This requires the redefinition:
$V_{C_i}\ra  G_{C_i\ov C_i}^{1/2}\ V_{C_i}$. 

For convenience, we shall discuss first the solution  with the total 
internal gauge flux $f^j\equiv \Fc^j$ turned off, \ie $b^j=-2\pi F^j$. 
According to \anglerelb, this means, that in the case of vanishing fluxes,
on the dual \ta side, the $D6$--branes become parallel $\th^j=0$. From 
\eqq \DN, in the case without fluxes $f^j\ra 0$, the mixing parameter becomes 
$\pm 1$, depending on the boundary conditions of the open string 
w.r.t. the plane $j$. In that case the system \dgl\ simplifies.

Let us first consider the matter fields $C^9_i$ from the $D9$--brane without fluxes.
In that case, all open string coordinates obey Neumann boundary conditions. Therefore we have
$D^j=+1$ and find:
\eqn\Dninematter{
G_{C^9_i\ov C^9_i} =\ap^{-3/2}\ e^{-\phi_4}\ 
\sqrt{\fc{(T^k-\ov T^k)(T^l-\ov T^l)}{T^i-\ov T^i}}\ \ \ ,\ \ \ 
(i,k,l)=\overline{(1,2,3)}\ .}
Now let us move on to the charged matter fields $C^{5,k}_i$ arising from a $D5$--brane, 
which is wrapped around the torus $T^{2,k}$. All open string coordinates orthogonal to the 
$D5$--brane obey Dirichlet boundary conditions, \ie $D^j=-1$ 
w.r.t. the two transverse tori $T^{2,j}$.
On the other hand, we have $D^k=+1$ w.r.t. $T^{2,k}$.
To be more concrete, let us discuss the case $k=1$, \ie the $D5$--brane is wrapped 
around the torus $T^{2,1}$. In that case, the equations \dgl\ are solved by:
\eqn\Dfivematter{
G_{C^{5,1}_i\ov C^{5,1}_i} =e^{-\phi_4} \times \cases{
\ap^{1/2}\ \fc{1}{\sqrt{(T^1-\ov T^1)(T^2-\ov T^2)(T^3-\ov T^3)}}\ ,  &$i=1\ ,$\cr
\ap^{-3/2}\ \sqrt{\fc{(T^1-\ov T^1)(T^2-\ov T^2)}{T^3-\ov T^3}}\ ,&$i=2\ ,$\cr
\ap^{-3/2}\ \sqrt{\fc{(T^1-\ov T^1)(T^3-\ov T^3)}{T^2-\ov T^2}}\ ,&$i=3$\ .}}
The other cases $k=2,3$ may be obtained from \Dfivematter\ by permutation of the planes.
So far, we have written our metrics in terms of the string moduli introduced in \stringmoduli\
to which the string vertex operators \vertexTU\ refer.
To obtain the metrics \Dninematter\ and \Dfivematter\ in the field theory basis, we only
have to replace the $T^j$--moduli through the field theory moduli $s$ and $t^j$, introduced
in \gaugeDnine\ and \gaugeDfive. This leads to
\eqn\untwistedmatter{\eqalign{
G_{C^9_i\ov C^9_i}&=\fc{\kappa_4^{-2}}{t^i+\ov t^i}\cr
G_{C^{5,1}_i\ov C^{5,1}_i}&\sim \cases{
\fc{\kappa_4^{-2}}{s+\ov s}\ ,  &$i=1\ ,$\cr
\fc{\kappa_4^{-2}}{t^3+\ov t^3}\ ,&$i=2\ ,$\cr
\fc{\kappa_4^{-2}}{t^2+\ov t^2}\ ,&$i=3$\ .}}}

In the following we allow for non--vanishing fluxes $f^j$ on the $D9$--brane.
With the same arguments as in the case of gauge couplings, we shall stick to
the $D9$--case only. The relevant calculation has been already performed in the
previous subsection. We only have to take into account the flux dependence of
$D^j$, given in \complexD\ and integrate the differential equations \dgl.
The solution is:
\eqn\matterflux{\eqalign{
G_{C_i\ov C_i}={\rm const.}\ e^{-\phi_4}\ \sqrt{\fc{T^i-\ov T^i}{(T^k-\ov T^k)(T^l-\ov T^l)}}\ 
&\fc{|T_2^k+if^k||T_2^l+if^l|}{|T_2^i+if^i|}\  ,}}
for $(i,k,l)=\overline{(1,2,3)}$.
It is evident, that this expression boils down to \Dninematter\ in the case of vanishing
fluxes $f^j=b^j+2\pi\ap F^j=0$.
In the case of N=1 supersymmetry, we may borrow \eqq \fluxcondition\ to cast \matterflux\ into 
\def\tf{\tilde{f}}
\eqn\matterfluxi{
G_{C_i\ov C_i}=\ap^{-3/2}\ e^{-\phi_4}\  \lf|1-\tf^k\tf^l\ri|\ 
\sqrt{\fc{(T^k-\ov T^k)(T^l-\ov T^l)}{T^i-\ov T^i}}\ \ \ ,\ \ \ 
(i,k,l)=\overline{(1,2,3)}\ ,}
where $\tf^i=f^i/T_2^i$. This 
should be compared with \Dninematter.
Finally, written in terms of the field--theory moduli $s$ and $t^j$, we obtain:
\eqn\matterfluxii{
G_{C_i\ov C_i}= \kappa_4^{-2}\lf|1-\tf^k\tf^l\ri|\ \fc{1}{t^i+\ov t^i}
\ \ \ ,\ \ \ (i,k,l)=\overline{(1,2,3)}\ .}
Note, that $\tf^i$ is a dimensionless quantity.
Through \duality\ the metric \matterfluxii\ translates to the \ta side with intersecting branes 
into:
\eqn\matterfluxiii{
G_{C_i\ov C_i}=\kappa_4^{-2}\lf|1-\tf^k\tf^l\ri|
\ \fc{1}{u^i+\ov u^i}\ \ \ ,\ \ \ (i,k,l)=\overline{(1,2,3)}\ ,}
where now $\tf^i=\tan(\pi\theta^i)$.

\subsec{Four--point amplitudes}

By internal charge conservation, the type $IIB$ 
three--point function with two matter
fields and one $U$ modulus vanishes and we instead proceed by 
factorizing a four--point function. 
The relation between the $\al'\to 0$ limit of the string four--point function
and the supersymmetric effective theory has been studied in \doubref\DKL\KL\
for the heterotic string. Although some details are different on the
type $I$ string side, the computation on the field theory side is the same and
we refer to the clear exposition in \DKL\ for details.  

The $\al'\to 0$ limit of the string correlation function
$\Ac=\ll C^a M^b \bb M^c \Cb^d \rr$ 
with two matter fields $C$ and two moduli fields $M$ is \DKL
\eqn\ftri{
\Ac\,\sim\,
G_{C^a\Cb^d}G_{M^b\bb M^c}\fc{ts}{u}+s\,R_{C^a\bb C^d M^b\bb M^c} +  \Ac_{pot}  +\cx O(k^4)\ ,}
where $R$ is the 
Riemann curvature of the K\"ahler manifold
$$
R_{C^a\bb C^d M^b\bb M^c}=
K_{C^aM^b\bb M^c\Cb^d}-K_{C^aM^b\Cb^e}\, G^{\Cb^eC^f}\, K_{\Cb^d \bb M^cC^f}
$$
and the kinematic invariants are defined in \mandelstamm.
The
$k_i$ are the momenta of the external fields in $\Ac$,
and $G_{C\Cb}=K_{C\Cb}\equiv \p_C\p_{\Cb}K$,
with $K$ the K\"ahler potential. The first term in \eqq \ftri\ 
is due to graviton exchange,
the second term from the sigma model couplings arising from the 
K\"ahler potential $K$ and the term $\Ac_{pot}$ denotes couplings due to 
the scalar potential from $F$- and $D$-terms. 
The first two terms in \ftri\ can be used to determine the moduli dependence
of the K\"ahler potential for the matter fields by integration.

For the present case of a compactification on the factorized tori
$T^{2,i}$, there are some significant simplifications. The metric for
the untwisted moduli fields is diagonal,
$G_{M^a\bb M^b}= \delta^{ab}(M^a+\ov M^a)^{-2}$. For $M$ a complex structure
modulus, $M=u^i$, the contribution $\Ac_{pot}$ vanishes and moreover the 
$u$ dependence of the the matter metric factorizes
\eqn\fmet{
G_{C^a\Cb^b}=\delta_{ab}\, \tx G_a \prod_i H_a^i(u^i).
}
Here $\tx G$ contains the dependence on all other moduli but the $u^i$.
For quadratic
K\"ahler potential the metric can be always diagonalized as above and
we will sometimes suppress the index $a$ for simplicity.
For further convenience we note, that for a simple dependence
$H(u)=(u+\ov u)^{-q},$ the ratio of the coefficients of the graviton 
and the sigma model channels in \ftri\ is the exponent $q$.

The $U$ moduli dependence of the matter metric may be extracted from the 4-point
function $\Ac=\langle V_{C^a}  V_{u^m} V_{\ov u^n} V_{\Cb^b} \rangle$,
$$
\Ac\,\sim\, \prod \int dx_i \, \cx J(x_i)\, \langle V^{-1}_{C^a}(x_1,k_1)\, 
V^0_{u^m}(x_2,k_2;x_3,k_3) \, 
V^0_{\ov u^n}(x_4,k_4;x_5,k_5)\,  V^{-1}_{\Cb^b}(x_6,k_6) \rangle,
$$
where $\cx J(x_i)$ is the Jacobian for fixing of the $PSL(2,\bf R)$ symmetry.

The only non-trivial point is the evaluation of the integrals over 
three real positions. 
We proceed by choosing a singular gauge choice for $PSL(2,\bf R)$, where
all bulk coordinates are proportional to a scale factor $w\in \bf R$, 
$
x_i=w\,y_i,\ i=2,...,5.
$ 
The Jacobian for the coordinate transformation
is singular at $w=0,\infty$ and the integrals need to be 
regularized by cutting small discs around these values.
The $w$-integral will eventually decouple. 
A particularly convenient choice is 
\eqn\gauge{
x_1=\infty,\ \ x_2=w\,y, \ \ x_3=w\,\yb, \ \ x_4=w,\ \ x_5=w,\ \ x_6=0.
}
At $w=0,\infty$  the bulk operators collide
with a boundary operator. Note that the bulk operators have no singularity as
they approach a generic point on 
the boundary and in fact the special choice \gauge\ places
the second bulk operator at the boundary.

It is now straightforward to evaluate the integrand, 
and we find
\eqn\risld{\eqalign{
\Ac&\sim \ \int \fc{dw}{w}\, \int d^2y\, |1-y|^{u-4}\,|y|^t\, 
\lf[ 2\ r_1\ (1-u/2) -\h r_2\ u \, (\fc{1-\yb}{\yb}+\fc{1-y}{y}) \ri]
\cr
&\sim \ r_1\, \fc{ts}{u} +r_2 \, s + \cx O(k^4) \ ,
}}
where
$r_1=\delta_{ab}\, \delta_{mn}$ and $r_2=r_1\, \delta_{am}$. Note
that the $w$-integral decouples, as promised.
The $y$ integral has the standard representation in terms of Gamma functions
$$
\int d^2y\, y^a \yb^{a'} (1-y)^b (1-\yb)^{b'} = \pi\,  
\fc{\Ga(1+a)\Ga(1+b)}{\Ga(2+a+b)}
\, \fc{\Ga(-1-a'-b')}{\Ga(-a')\Ga(-b')},
$$
(provided $a-a'\in \IZ$,\ $b-b'\in \IZ$)
and leads to the small momentum expansion quoted above.

As noted below \fmet\ the relative coefficient of the two terms 
at $\cx O(k^2)$ determine the $u$ dependence
of $G_{C^a\Cb^a}$ to be 
\eqn\utudep{\prod_i H_a^i(u^i)=(u^a+\ov u^a)^{-1}.}

Let us summarize  at this point all 
the above results in terms of the 
K\"ahler potential for the untwisted type $IIB$ moduli and to second
order in the matter fields:
\eqn\puttogether{\eqalign{\cr
\kappa_4^2\, 
K=&-\ \ln(s+\ov s)-\sum_{i=1}^3\ln(t^i+\ov t^i)-\sum_{i=1}^3\ln(u^i+\ov u^i)\cr
&+\sum_{i=1}^3 \fc{|C_i^9|^2}{(u^i+\ov u^i)(t^i+\ov t^i)}
\, \sqrt{\fc{(1+(\tf^k)^2)(1+(\tf^l)^2)}{1+(\tf^i)^2}}\cr
&+\sum_{i=1}^3\fc{|C_i^{5,i}|^2}{(s+\ov s)(u^i+\ov u^i)}
+\sum_{i,j,k}^3\fc{d_{ijk}\ |C_j^{5,k}|^2}{(t^i+\ov t^i)(u^j+\ov u^j)}.\cr\ }}
Here we have introduced the tensor $d_{ijk}$ which is 1 for $(i,j,k)$ a 
permutation of $(1,2,3)$ and 0 otherwise. Note that \puttogether\ is written
in the Einstein frame, which accounts for an extra factor $e^{2\phi_4}$ 
in the metric.
In the case of fixed complex structure and zero fluxes $\tf$, 
the above K\"ahler potential agrees with the results of \AFIV, 
obtained there by $T$--duality arguments. 

\newsec{Twisted matter metric}
\subsec{Twist fields and twist field correlators}
In the following we will work in type $IIA$. Here,
in intersecting brane world models, the chiral matter comes from open strings stretched between
two intersecting branes. Hence these open strings obey mixed boundary conditions \BDL:
\eqn\boundary{\eqalign{
\lf.\re\ \p_\sigma Z^j\ri|_{\sigma=0}
\ \ ,&\ \ \lf.\im\ Z^j\ri|_{\sigma=0}=0\ ,\cr
\lf.\re\ e^{i\th^j} \p_\sigma Z^j\ri|_{\sigma=\pi}
\ \ ,&\ \ \lf.\im\ e^{i\th^j} Z^j\ri|_{\sigma=\pi}=0\ ,}}
with $\th^j$ being the relative 
angle of the two branes w.r.t. the $j$--th internal torus $T^{2,j}$.
These boundary conditions produce cuts $\sim z^{-\th^j}$
in the map $z\ra Z^j(z)$ from the disk to the target space. 
In the vertex operator for the corresponding scalar matter field $C_\th$, these cuts 
are introduced\foot{This is notorious from heterotic orbifold compactifications \DFMS.} 
on the disk by the twist fields $\sigma_{\th^j}(z)$.
The world sheet 
supercurrent on the boundary ($z=\ov z$) of a disk is given by:
\eqn\supercurrent{\eqalign{
T_F(z,\ov z)&=\h\ [T_F(z)+\tilde T_F(\ov z)]\ ,\cr
T_F(z)&=\p X_\mu(z)\psi^\mu(z)+\sum_{i=1}^3\ \p \ov Z^i(z)\Psi^i(z)+\p Z^i(z)\ov\Psi^i(z)\ ,\cr
\tilde T_F(\ov z)&=\ov\p X_\mu(\ov z)\tilde \psi^\mu(\ov z)+\sum_{i=1}^3\ 
\ov\p \ov Z^i(\ov z)\tilde
\Psi^i(\ov z)+\ov \p Z^i(\ov z)\ov\Psi^i(\ov z)\ .}}
The latter is only invariant under the twist if the internal fermions 
$\Psi,\tilde\Psi$, which may be bosonized according to 
$\Psi^j(z)=e^{iH^j(z)},\tilde \Psi^j(\ov z)=e^{i\tilde H^j(\ov z)}$ 
are twisted as well. This is accomplished 
by the fermionic twist fields $s_{\pm\th^j}(z)$, which are bosonized 
$s_{\pm\th^j}(z)=e^{\pm i(1-\th^j)H^j(z)}$.  
Here, the fields $H^j(z)$ correspond to the internal bosonized world--sheet fermion
($j=1,2,3$).

Generically two intersecting stacks $a$ and $b$ (of $N$ and $M$ $D6$ branes, respectively)
establish\foot{In the case, when both stacks 
do not sit on top of the  orientifold planes. Otherwise, if one stack is 
placed at an orientifold plane the gauge group is $SO(2N)$ or $USp(N)$.} the 
gauge group $U(N)\times U(M)$.
The massless (twisted) $R$--sector of open strings stretched between these two stacks $a$ and $b$
intersecting at an angle $\th$
gives rise to massless chiral fermions in the bifundamental $(N,\ov M)$
of $U(N)\times U(M)$.
In the case of N=1 supersymmetry these fermions build chiral N=1 multiplets 
whose scalars (scalar matter fields $C_\th,\ \ov C_\th$) stem from the open strings in the 
(twisted) $NS$--sector.  
Hence these chiral multiplets sit at the intersection of the two stacks $a$
and $b$ and their vertex operators have to be
inserted at the boundary of any string world sheet diagram.
The vertex operator for an N=1 matter field $C_\th$ inserted at the boundary $z=\ov z$ of the disk
is (\cf \eg \BDL)
\eqn\mattervertex{
V_{C_\th}^{(-1)}(z,k)=\lambda\ 
e^{-\phi(z)}\ \prod_{j=1}^3 s_{\th^j}(z)\ \sigma_{-\th^j}(z)\ e^{ik_\nu X^\nu(z)}\ }
in the $-1$ ghost picture. 
Here, the twist fields $s_{\th^j},\sigma_{\th^j}$ 
depend on the angles $\th=(\th^1,\th^2,\th^3)$
between the two intersecting branes describing the details of the chiral matter 
under consideration. The bosonic twist fields $\sigma_{\th^j}$ are responsible for twisting the
$NS$ ground state, while the twisted spin fields $s_{\th^j}$ are required by supersymmetry
(see the argument after \eqq \supercurrent).
The Chan--Paton factor $\lambda$ describes the gauge degrees of freedom at both string ends.
Similarly, for the conjugate matter field $\ov C_\th$, originating from the same open string,
but with different orientation, we have:
\eqn\mattervertexconj{
V_{\ov C_\th}^{(-1)}(z,k)=\lambda^\dagger\ 
e^{-\phi(z)}\ \prod_{j=1}^3 s_{-\th^j}(z)\ \sigma_{\th^j}(z)\ e^{ik_\nu X^\nu(z)}\ .}

The techniques of correlation functions with twist field operators $\sigma_{\th^j}$
on the sphere have been
pioneered in \doubref\DFMS\HV\ and generalized  in \Fourref\BKM\EJSS\casas\koba. 
These results may be borrowed to also determine twist field  correlation functions on the disk
\doubref\GNS\benakli.
Correlators with twist fields located on the boundary $z=\ov z$ of a disk 
are essentially obtained\foot{This is not true for scattering processes involving 
both fields from the boundary {\it and} closed strings from the bulk.}
from the corresponding results on the sphere by ``taking the square root''. 
This procedure leads to the following basic correlators on the disk (for one complex 
twisted dimension):
\eqn\basic{\eqalign{
\vev{\sigma_\th(z_1)\ \sigma_{-\th}(z_2)}&=(z_1-z_2)^{-\th(1-\th)}\ ,\cr
\vev{s_\th(z_1)\ s_{-\th}(z_2)}&=(z_1-z_2)^{-(1-\th)^2}\ .}}
The fields $\sigma_{\th}(z)$ and $s_{\th}(z)$ have conformal dimensions 
$h_{\sigma_\th}=\h \th(1-\th)$ and $h_{s_\th}=\h(1-\th)^2$, respectively.
With this information it is easy to realize that the vertex operator $V_{C_\th}^{(-1)}(z,k)$ 
has conformal weight $h=1$, provided one uses the supersymmetry relation 
in the form $\sum_i \th^i=2$.
The twist fields generate branchings on the complex fields $\p Z(z),\ov\p Z(\ov z), 
\Psi(z),\tilde \Psi(\ov z)$, introduced in \complexify. The local behavior
of those fields in the presence of twist fields is given by the operator 
products \doubref\DFMS\david
\eqn\ope{\hskip-0.3cm\eqalign{
\p Z^j(z)\ \sigma_{\th^j}(w)       =(z-w)^{\th^j-1} \tau_{\th^j}(w)+\ldots ,&\  
\p Z^j(z)\ \sigma_{-\th^j}(w)=(z-w)^{-\th^j} \tilde\tau_{-\th^j}(w)+\ldots,\cr
\p \ov Z^j(z)\ \sigma_{\th^j}(w)   =(z-w)^{-\th^j} \tilde\tau_{\th^j}(w)+\ldots ,&\ 
\p \ov Z^j(z)\ \sigma_{-\th^j}(w)  =(z-w)^{\th^j-1} \tau_{-\th^j}(w)+\ldots,\cr
\ov\p Z^j(\ov z)\ \sigma_{\th^j}(w)=-(\ov z-w)^{\th^j-1} \tau_{\th^j}(w)+\ldots ,&\   
\ov\p Z^j(\ov z)\ \sigma_{-\th^j}(w) =-(\ov z-w)^{-\th^j} \tilde\tau_{-\th^j}(w)+\dots,\cr
\ov\p\ov Z^j(\ov z)\ \sigma_{\th^j}(w)=-(\ov z-w)^{-\th^j} \tilde\tau_{\th^j}(w)+\dots ,&\  
\ov\p\ov Z^j(\ov z)\ \sigma_{-\th^j}(w)=-(\ov z-w)^{\th^j-1}\tau_{-\th^j}(w)+\dots.}}
Obviously, the OPEs involving negative angles $\-\th^j$ are obtained from the ones with positive 
angles $\th^j$ after performing the replacement $\th^j\ra 1-\th^j$.
In addition, we have:
\eqn\opei{\eqalign{
\Psi^j(z)\ s_{\th^j}(w)&=(z-w)^{1-\th^j}\ \tilde t_{\th^j}(z)+\ldots\ ,\cr
\Psi^j(z)\ s_{-\th^j}(w)&=(z-w)^{\th^j-1}\ t_{-\th^j}(z)+\ldots\ ,\cr
\ov \Psi^j(z)\ s_{\th^j}(w)&=(z-w)^{\th^j-1}\ t_{\th^j}(z)+\ldots\ .}}
After applying the picture changing operator $P(w,\ov w)=e^{\phi(w)}T_F(w)+e^{\tilde \phi(\ov w)}
\tilde T_F(\ov w)$, we obtain the matter field vertex in the zero ghost picture:
\eqn\mattervertexi{
V_{C_\th}^{(0)}(z,k)=\lambda\ \sum_{l=1}^3\lf[t_{\th_l}(z)\tau_{-\th_l}(z)
 +\fc{1}{3}\ i\,(k_\mu\psi^\mu)\ 
s_{\th_l}(z)\sigma_{-\th_l}(z)\ri]\prod_{j=1\atop j\neq l}^3s_{\th^j}(z)\sigma_{-\th^j}(z)
\ e^{ik_\nu X^\nu(z)}\ .}
Here, $\tau_{\th^j}(z)$ are excited bosonic twist fields of conformal 
dimension $h_{\tau_\th}=\h (1-\th)(2+\th)$.
On the other hand, the excited fermionic twist fields $t_{\th^j}(z)$, which may be bosonized
according to $t_{\pm\th^j}(z)=e^{\mp i\th^j H^j(z)}$, have 
conformal dimension $h_{t_\th}=\h\th^2$.
With this information, the operator \mattervertexi\ 
has conformal weight $h=1$ after imposing the condition \susy.
In the supersymmetric case, the matter field vertex operator represents a marginal 
deformation of the underlying N=2 superconformal field theory.
This is why it has to carry conformal weight $h=1$ in that case and 
the matter field is a massless scalar.
In fact, in the supersymmetric case, the scalar matter field vertex  $V_{C_\th}^{(0)}$ 
is the highest component of an N=2 superconformal multiplet, 
whose fermionic component is represented by the fermionic vertex operator.

 From the local behavior of the twist fields, given in \ope, we may derive
the  following correlators, which we shall need later:
\eqn\needcorr{\eqalign{
\vev{\p \ov Z^j(z)\ \p Z^j(w)\ \sigma_{\th^j}(x_1)\ \sigma_{-\th^j}(x_2)}&=
-\fc{(x_1-x_2)^{-\th^j(1-\th^j)}}{(z-w)^2}\ \fc{(w-x_1)^{\th^j-1}\ (z-x_2)^{\th^j-1}}{
(w-x_2)^{\th^j}\ (z-x_1)^{\th^j}}\ ,\cr
&\times [(z-x_2)(w - x_1) +\th^j (z-w)(x_1 - x_2)  ]\ ,\cr
\vev{\ov\p Z^j(\ov z)\ \p \ov Z^j(w)\ \sigma_{\th^j}(x_1)\ \sigma_{-\th^j}(x_2)}&=
-\ov{D^{j}}\ \fc{(x_1-x_2)^{-\th^j(1-\th^j)}}{(\ov z-w)^2}\ 
\fc{(\ov z-x_1)^{\th^j-1}\ (w -x_2)^{\th^j-1}} {(\ov z-x_2)^{\th^j}\ (w-x_1)^{\th^j}}\cr
&\times  [(w - x_2)(\ov z-x_1)  +\th^j(w-\ov z)(x_1-x_2)]\ ,\cr
\vev{\ov \p \ov Z^j(\ov z)\ \p Z^j(w)\ \sigma_{\th^j}(x_1)\ \sigma_{-\th^j}(x_2)}&=
-D^{j}\ \fc{(x_1-x_2)^{-\th^j(1-\th^j)}}{(\ov z-w)^2}\ 
\fc{(w-x_1)^{\th^j-1}\ (\ov z-x_2)^{\th^j-1}}{
(w-x_2)^{\th^j}\ (\ov z-x_1)^{\th^j}}\cr
&\times  [ (w - x_1) (\ov z-x_2)+ \th^j(\ov z-w)(x_1-x_2)]\ .}}
The contraction of internal fermionic fields with spin fields referring to the same 
complex plane becomes on the disk:
\eqn\needcorri{\eqalign{
\vev{\tilde\Psi^j(\ov z)\ \ov{\Psi^j}(w)\ s_{\th^j}(x_1)\ s_{-\th^j}(x_2)}&=
\fc{\ov{D^j}}{\ov z-w}\ (x_1-x_2)^{-(1-\th^j)^2}\ 
\lf[\fc{(\ov z-x_1)(w-x_2)}{(\ov z-x_2)(w-x_1)}\ri]^{1-\th^j}\ \ ,\cr
\vev{\ov{\tilde\Psi^j}(\ov z)\ \Psi^j(w)\ s_{\th^j}(x_1)\ s_{-\th^j}(x_2)}&=
\fc{D^j}{\ov z-w}\ (x_1-x_2)^{-(1-\th^j)^2}\  
\lf[\fc{(\ov z-x_2)\ (w-x_1)}{(\ov z-x_1)\ (w-x_2)}\ri]^{1-\th^j}\ .}}

Finally, the two--point correlators of excited twist on the disk fields may be obtained 
by taking the square root of the closed string results \DFMS\ (\cf also \david):
\eqn\basici{\eqalign{
\vev{\tau_\th(z_1)\ \tau_{-\th}(z_2)}&=\th\ (z_1-z_2)^{-\th(3-\th)}\ ,\cr
\vev{t_\th(z_1)\ t_{-\th}(z_2)}&=(z_1-z_2)^{-\th^2}\ .}}
On the other hand, these correlators may be also derived from \eqqs \ope\ and \needcorr.

\subsec{String computation of moduli dependence}
Let us now calculate the string $S$--matrix\foot{In this section we are in the type $IIA$
picture of intersecting $D6$--branes, where we however omit the prime on the
moduli fields $U^j$.} with
\eqn\todo{
\Ac_{\A\oA U^j}=\fc{i}{2\pi}\, \int \fc{dz_1 dz_2 d^2z_3}{V_{\rm CKG}}\ 
\vev{V_\A^{(-1)}(z_1,k_1)\ V_{\oA}^{(-1)}(z_2,k_2)\ V_{U^j}^{(0,0)}(\ov z_3,z_3,q)}}
on the disk to extract the information on the metric $G_{C_\th\ov C_\th}$ 
of the two matter fields $C_\th,\ov C_\th$.
The two matter field vertices \mattervertex\ and \mattervertexconj\ 
are inserted at the boundary of the disk, \ie $z_1=\ov z_1,z_2=\ov z_2$. 

The correlation functions such as \needcorr\ are in fact the standard
correlators for the Neumann boundary conditions and not the right ones
for doing the computation for branes at angles. However we know already
from the  discussion in section 3.4. that there is a simple way to 
adapt to this case: it suffices to complex conjugate the right-moving fields.
Thus we write the closed string vertex operator for the complex structure
modulus $U^j$ as 
\eqn\vertexU{
V_{{U}^j}^{(0,0)}(\ov z,z,k)=\fc{1}{{U}^j-\ov {U}^j}\ \lf[\ov \p Z^j+i(k\tilde \psi)\ 
\tilde{\Psi^j}(\ov z)\ri]\ 
\lf[\p \ov Z^j+i(k\psi)\ {\ov \Psi^j}(z)\ri]\ e^{iq_\nu X^\nu(\ov z,z)},}
and subsequently use the correlators above. To compute the
correlation function of the physical scalar we will again consider
the sum of the above amplitude and the one with the $U$ operator
replaced by the complex conjugate at the end.

There are two non--vanishing contributions to the amplitude \todo.  They
arise from the contraction of the matter vertices with either both bosons or both fermions of 
the $U^j$--vertex operator. We shall denote these two possibilities by $X_1$ and $X_2$, 
respectively. With the vertex operator \vertexU, 
the relevant correlators accounting for the internal sector are given in 
\needcorr and \needcorri.
The correlator for the exponentials is given by the correlator $\Ec$ introduced  in \expo.
The contractions of the space--time fields are basic and given in \green.
After putting all contributions together and  using the supersymmetry condition \susy,
the amplitude \todo\ becomes
\eqn\Todo{
\Ac_{\A\oA U^j}=\fc{1}{\pi i}\, 
\fc{\ov{D^j}}{U^j-\ov U^j}\ \int \fc{dz_1 dz_2 d^2z_3}{V_{\rm CKG}}\ \Ec\ 
\ (z_1-z_2)^{-2}\ (\ov z_3-z_3)^{-2}\ (X_1-2\ t\  X_2)\ ,}
with:
\eqn\zusatz{\eqalign{
X_1&=-\fc{(\ov z_3-z_1)^{-\th^j}\ (z_3-z_2)^{-\th^j}}{(\ov z_3-z_2)^{1-\th^j}\ 
(z_3-z_1)^{1-\th^j}}\ \lf[(z_3 - z_2)(\ov z_3-z_1)  +(1-\th^j)(z_3-\ov z_3)(z_1-z_2)\ri]\ ,\cr 
X_2&=\lf[\fc{(\ov z_3-z_1)(z_3-z_2)}{(\ov z_3-z_2)(z_3-z_1)}\ri]^{1-\th^j}\ .}}
As in the case with the untwisted matter field correlator, we fix three vertex positions
according to \position\ and include the $c$--ghost correlator \cghost.
To this end, the string $S$--matrix $\Ac_{\A\oA U^j}$ may be expressed in terms of the integrals
\integralii:
\eqn\TTodo{\eqalign{
\Ac_{\A\oA U^j}&=-\fc{i\ov{D^j}}{2\pi U_2^j}\, 2^{-4t-3} 
\lf[(1+2t) I(-1-2t,2\th^j-2)-4i (1-\th^j) I(-2t,2\th^j-1)\ri]\cr
&=-\fc{i\ov D^j}{2U^j_2}\ e^{-i\pi\th^j}\ 
\fc{\Gamma(-2t)\ \lf[\ \Gamma(-t+\th^j)+(1-\th^j)\ \Gamma(-1-t+\th^j)\ \ri]}
{\Gamma(-1-t+\th^j)\ \Gamma(-t+\th^j)\ \Gamma(1-t-\th^j)}\ .}}
The above expression has the following momentum expansion:
\eqn\momentexp{
\Ac_{\A\oA {U}^j}=-\fc{i\ov D^j}{4\pi U^j_2}\ e^{-i\pi\th^j}\ 
\sin(\pi \theta^j)\, 
\lf[\ 1+t\ \rho^j +\Oc(t^2)\ \ri]\ ,}
where $\psi(x)=\fc{d}{dx}\ \ln\Gamma(x)$ and
$$
\rho^j=2\gamma_E+\psi(\th^j)+\psi(1-\th^j)\ .
$$
Using  $D=e^{-2i\pi\th^j}$, the symmetrized amplitude is finally
\eqn\momentexpi{
\Ac_{\A\oA {U_2}^j}=\fc{1}{2\pi U^j_2}\ \cos(\pi \theta^j)\,
\sin(\pi \theta^j)\, 
\lf[\ 1+t\ \rho^j +\Oc(t^2)\ \ri]\ .}
The linear term of the $t$-expansion \momentexpi\ describes the derivative of
the matter fields metric $G_{C_\th\ov C_\th}$ w.r.t. $U_2^j$. After redefining the vertex 
operators $V_{C_\th}\ra G_{C_\th\ov C_\th}^{1/2}\ V_{C_\th}$, 
which converts the metric in the canonical normalized string 
basis into the field--theory basis, we obtain the 
following differential equation
\eqn\august{
\fc{\p}{\p U_2^j}\ G_{C_{\th}\ov C_\th}= -G_{C_{\th}\ov C_\th}\ 
\fc{1}{2\pi U^j_2}\ \cos(\pi \theta^j)\,
\sin(\pi \theta^j)\, \rho^j.}
Obviously, for all three complex planes $j=1,2,3$, the scattering amplitude \todo\ yields 
the same type of equation.
Through \anglerel\ the angle $\th^j$ depends on the moduli $U^i$. From 
\anglerelb\ we obtain 
\eqn\anglerelation{
\fc{\p \th^j}{\p U_2^j}=-\fc{1}{\pi U^j_2}\cos(\pi\theta^j)\sin(\pi\theta^j).}
With these relations, we may convert the equations 
\august\ into differential equations w.r.t. $\th^j$:
\eqn\augusti{
\fc{\p}{\p \th^j}\ \ln G_{C_{\th}\ov C_\th}= \fc{\rho^j}{2}.}
These equations integrate to:
\eqn\finally{
G_{C_{\th}\ov C_\th}\sim e^{\gamma_E\ \sum\limits_{j=1}^3\th^j }\ \prod_{j=1}^3
\sqrt{\fc{\Gamma(\th^j)}{\Gamma(1-\th^j)}}\ .}
Note, that this expression depends on the complex structure moduli $U^j$ 
through the 
relation \anglerel. 

The dependence of the metric $G_{C_\th\Cb_{\th}}$ on the moduli $t^i$
can be computed from the four--point disk diagram
in the twisted sector, similarly as in the untwisted case.
The relevant amplitude $
\Ac=\ll V_{C_\th}  V_{t^m} V_{\ov t^m} V_{\Cb_{\th}}\rr$ is:
\eqn\ampii{
\Ac \sim \prod \int dx_i \, \cx J(x_i)\, \ll V^{-1}_{C_\th}(k_1,x_1)\, 
V^0_{t^m}(k_2,x_2;k_3,x_3) \, 
V^0_{\bb t^m}(k_4,x_4;k_5,x_5)\,  V^{-1}_{\Cb_{\th}}(k_6,x_6) \rr.
}
We proceed exactly  as in the untwisted case and find
\eqn\risldii{\eqalign{
\Ac&\sim\ \int \fc{dw}{w}\, \int d^2y\, |1-y|^{u-4}\,|y|^{t-2\th'}\, 
\lf\{\ F(1,y)F(1,\bb y)-\fc{u}{4}[F(1,y)+F(1,\bb y)]\ \ri\}\cr
&\sim \  \fc{ts}{u} +(1-\th')\, s + \cx O(k^4) \ ,
}}
where $F(a,b)=(1-\th')+\th'\fc{b}{a}$ and $\th'=1-\th^m$. 
The relative coefficients of the two terms in the last line of the
above equation determine the moduli dependence of the metric:
\eqn\metmet{
G_{C_\th\ov C_\th}=\tx G\ \prod_m (t^m+\ov t^m)^{-\th^m}.
}
Here $\tx G$ an arbitrary function of the other fields.

\newsec{Scattering of four matter fields and normalization of Yukawa couplings}
In the following we study 4-point functions of four matter fields 
\eqn\ampfm{
\Ac_{ij}=C_A\, 
\cdot \langle V_{C_i}(z_1)V_{\ov C_i} (z_2)V_{C_j}(z_3)V_{\ov C_j}(z_4)\rangle, 
} 
to obtain further information on the normalization of the 
physical Yukawa couplings and the 
holomorphic superpotential of the matter fields. 
Here $C_i$ and $C_j$ are again brane matter fields and $C_A$ an yet undetermined
normalization constant.

If all matter fields are  charged under a common gauge group, the leading 
term in an momentum expansion of the factorization limit $z_1\to z_2$ 
contains an $s$--pole $\sim \kappa =(t-u)/s$ from an exchange of a gauge
boson. For appropriate normalization this exchange agrees with the same 
process in the effective field theory
\eqn\fixC{
\Ac^\kappa_{ij}=\lim_{z_2\to z_1} \Ac_{ij}|_\kappa=  
g_A^2\ G_{C_i\ov C_i}\ G_{C_j\ov C_j},}
where we omitted identical group theory and kinematic factors 
on both sides. The subscript $|_\kappa$ denotes the
coefficient of the kinematics of the gauge boson exchange.
Moreover,  $g_A$ is the gauge coupling of the exchanged gauge boson on the brane $A(i)$
on which the matter field $C_i$ lives. 
The above equation may be used to eliminate the constant
$C_A$ in favor of the field theory quantities. 

If the string amplitude has been normalized as in \fixC,
the leading term in the factorization limit $z_3\to z_4$
reproduces the contact term from  the Yukawa couplings in the 
effective field theory
\eqn\fixY{
\Ac^Y_{ij}=\lim_{z_2\to z_4} \Ac_{ij}=  
\sum_k\, Y_{ijk}\ G^{C_k\ov C_k}\ \ov Y_{\ov k \ov j \ov i}=
e^{\kappa_4^2 \hat{K}}\sum_k W_{ijk}\ G^{C_k\ov C_k}\ 
\ov{W}_{\ov k \ov j \ov i}.
}
Here we used the relation between the physical Yukawa couplings and the 
holomorphic superpotential, at cubic order in the matter fields \CremmerEN\DKL,
\eqn\yukawa{
Y_{\alpha\beta\gamma}=
e^{\h\kappa_4^2\hat{K}}\ W_{\al\be\gamma}.}
If $W_{ijk}$ is ``invertible'' in the sense that one can chose the indices 
such that the sum over $k$ collapses, the two equations imply 
\eqn\fixW{
|W_{ijk}|^2=
e^{-\kappa_4^2 \hat{K}}\, g_A^2 \, 
G_{C_i\ov C_i}G_{C_j\ov C_j}G_{C_k\ov C_k}
\ \lf(\fc{A^Y_{ij}}{A^\kappa_{ij}}\ri)\ ,}
which is independent of the normalization constant.
The above equation relates the 
holomorphic superpotential and the Yukawa couplings in the
string frame determined by $A^Y_{ij}$.

Note that $W$ is holomorphic by definition  and on general
grounds we expect that for the, say,  type $IIA$ theory with branes at angles,
it depends only on the K\"ahler moduli $t^i$ \Doug. In this way the
two factorization limits discussed above lead to a 
non-trivial constraint on the expressions appearing on the r.h.s. 
For the K\"ahler potential \kpstu, the exponential factor becomes
$$
e^{-\kappa_4^2 \hat{K}}=e^{-4\phi_4}\, \prod_i U^{IIB,i}_2
=e^{-4\phi_4}\, \prod_i T^{IIA,i}_2,
$$
and is in fact independent of the geometric moduli $T$ ($U$) in the 
type $IIB$ ($IIA$) theory.

\subsec{Untwisted matter fields  $C_i$}

In the following we shall determine the scattering of four untwisted matter fields:
\eqn\todoo{\eqalign{
\Ac_{\ov C_i C_i\ov C_j C_j}(k_1,k_2,k_3,k_4)
&=C\ \int\limits^\infty_{-\infty}\ \prod\limits_{i=1}^4 dz_i\ V_{\rm CKG}^{-1}\cr  
&\times \vev{V_{\ov C_i}^{(0)}(z_1,k_1)\ V_{C_i}^{(-1)}(z_2,k_2)\ V_{\ov C_j}^{(0)}(z_3,k_3)\ 
V_{C_j}^{(-1)}(z_4,k_4)}\ .}}
Two matter fields $C_i,\ov C_i$ refer to the $i$--th subplane, whereas the other two fields 
$C_j,\ov C_j$ originate from the $j$--th subplane.
The constant $C$ is the normalization of the four--point function to be determined later.
All matter field vertex operators are inserted at the boundary of the disk.
In the zero ghost picture they have been given 
in \zeromatter, and in the $-1$ ghost picture they read:
\eqn\untwA{
V^{(-1)}_{C_i}(z,k)=\lambda\ e^{-\phi(z)}\ \Psi^i(z)\ e^{ik_\nu X^\nu(z)}\ .}
Due to internal charge conservation, a non--vanishing contribution arises in \todoo\ only if all
internal complex fermions are contracted. Therefore, we obtain:
\eqn\obtains{\eqalign{
 \Ac_{\ov C_i C_i\ov C_j C_j}(k_i)
&=-t\ C\int\ \prod\limits_{i=1}^4 dz_i\ V_{\rm CKG}^{-1}\ \Ec\ \cr  
&\times \fc{1}{z_2-z_4}\ \fc{1}{z_1-z_3}\ \lf[\fc{1}{(z_1-z_2)\ (z_3-z_4)}
+\fc{\delta^{ij}}{(z_2-z_3)\ (z_1-z_4)}\ri]\ .}}
The correlator $\Ec$ for the exponentials is given in \expo\ and is subject to the
momentum constraint \conserve.
We may use $PSL(2,\IR)$--invariance on the disk to fix three vertex positions
\eqn\positionfour{
z_1=0\ \ ,\ \ z_2=x\ \ ,\ \ z_3=1\ \ ,\ \ z_4:=z_\infty=\infty\ ,}
and introduce the factor $z_\infty^2$ to account for the $c$--ghosts.
In that case the expression \obtains\ reduces to
\eqn\reduce{\eqalign{
\Ac_{\ov C_i C_i\ov C_j C_j}(k_i)
&=-t\ C\ \Tr(\lambda_i^\dagger\lambda_i\lambda_j^\dagger\lambda_j)\ 
\int^1_{0}\ dx\ x^s\ (1-x)^u
\lf(  \fc{1}{x}+\delta^{ij}\ \fc{1}{1-x}  \ri)\cr
&-t\ C\ \Tr(\lambda_i^\dagger\lambda_j^\dagger\lambda_i\lambda_j)\ 
\int^1_{0}\ dx\ x^{t-1}\ (1-x)^u
\lf(1+\delta^{ij}\fc{1}{x-1}  \ri)\cr
&-t\ C\ \Tr(\lambda_i^\dagger\lambda_j^\dagger\lambda_j\lambda_i)\ 
\int^1_{0}\ dx\ x^{t-1}\ (1-x)^s
\lf(  \fc{1}{x-1}+\delta^{ij} \ri)}}
after taking into account the ordering of the Chan--Paton factors w.r.t. the vertex 
operator positions.
With the Veneziano integral 
\eqn\veneziano{
V(s,u):=\int_0^1\ dx\ x^s\ (1-x)^u=\fc{\Gamma(s+1)\ \Gamma(u+1)}{\Gamma(2+s+u)}\ ,}
we obtain:
\eqn\Obtain{\eqalign{
\Ac_{\ov C_i C_i\ov C_j C_j}(k_i)&=
-t\ C\ \Tr(\lambda_i^\dagger\lambda_i\lambda_j^\dagger\lambda_j)\ 
\lf[V(s-1,u)+\delta^{ij}\ V(s,u-1)\ri]\cr
&-t\ C\ \Tr(\lambda_i^\dagger\lambda_j^\dagger\lambda_i\lambda_j)\ 
\lf[V(t-1,u)-\delta^{ij}\ V(t-1,u-1)\ri]\cr
&-t\ C\ \Tr(\lambda_i^\dagger\lambda_j^\dagger\lambda_j\lambda_i)\ 
\lf[-V(t-1,s-1)+\delta^{ij}\ V(t-1,s)\ri]\ .}}
Adding the group factors from the flipped diagrams and 
expanding up to fourth order in the momenta we finally arrive at 
\eqn\alltogether{\eqalign{
\Ac_{\ov C_i C_i\ov C_j C_j}(k_i)
&=C\lf[
 \Tr(\lambda_i^\dagger\lambda_i\lambda_j^\dagger\lambda_j)+
 \Tr(\lambda_i^\dagger\lambda_j\lambda_j^\dagger\lambda_i)\ri]
\lf[\lf(-\fc{t}{s}+\fc{\pi^2}{6}tu\ri)+\delta^{ij}\lf(-\fc{t}{u}+\fc{\pi^2}{6}st\ri)\ri]\cr
&+C\lf[ \Tr(\lambda_i^\dagger\lambda_j^\dagger\lambda_i\lambda_j)+
 \Tr(\lambda_i^\dagger\lambda_j\lambda_i\lambda_j^\dagger)\ri]
\lf[\lf(-1+\fc{\pi^2}{6}tu\ri)+\delta^{ij}\lf(-\fc{s}{u}+\fc{\pi^2}{6}st\ri)\ri]\cr
&+C\lf[ \Tr(\lambda_i^\dagger\lambda_j^\dagger\lambda_j\lambda_i)+
 \Tr(\lambda_i^\dagger\lambda_i\lambda_j\lambda_j^\dagger)\ri]
\lf[\lf(-\fc{u}{s}+\fc{\pi^2}{6}tu\ri)+\delta^{ij}\lf(-1+\fc{\pi^2}{6}st\ri)\ri]\ .}}
The two factorization limits discussed above can now 
be read off from the general result \alltogether with $i\neq j$.
More precisely, the above result implies that the ratio $A^Y_{ij}/A^\kappa_{ij}$
is a constant, whereas it does not determine the index structure of 
the cubic couplings $W_{ijk}$. The latter can be inferred from 
the charge selection rules of the internal CFT, with 
the result that $W\neq 0$ only if $i\neq j\neq k$ \BL. Using the result 
\puttogether\ for the matter metric, and \susygaugenine\ for 
the gauge coupling,
one finds that the r.h.s. of \fixW\ is a constant, 
leading to a moduli independent superpotential
$$W={\rm const.}\ C^9_1C^9_2C^9_3.$$ A similar analysis can be performed for 
the other terms in the superpotential of \BL, containing various couplings  
for 5- and 9-brane matter fields, and again one finds that the
cubic terms in the superpotential are moduli independent. This is in 
contrast to formulae in the literature, where the cubic terms in $W$ 
sometimes appear with extra factors of the brane gauge couplings.

\subsec{Twisted matter fields $C_\th$}

Finally we consider the scattering of four twisted matter fields:
\eqn\todoi{\eqalign{
\Ac_{\A\oA\AA\oAA}(k_i)
&=C\int\ \prod\limits_{i=1}^4 dz_i\ V_{\rm CKG}^{-1}\cr  
&\times \vev{V_{\A}^{(0)}(z_1,k_1)\ V_{\oA}^{(-1)}(z_2,k_2)\ V_{\AA}^{(0)}(z_3,k_3)\ 
V_{\oAA}^{(-1)}(z_4,k_4)}\ ,}}
All matter field vertex operators are inserted at the boundary of the disk.
The above choice of matter field vertices \mattervertex\ and \mattervertexi\ in the 
$-1$ and $0$ ghost picture, respectively
has the advantage, that no contribution of excited twist or
spin fields contribute due to internal $U(1)$  charge conservation.
Hence from both vertex operators in the zero ghost picture \mattervertexi\
only the second term contributes. We immediately obtain the 
factor $t\ (z_1-z_3)^{-1}$ from contracting the two space--time fermions 
$\psi(z_1)$ and $\psi(z_3)$.
The contribution $\Ec$ of the exponentials is given in \expo, 
the ghosts give a factor
$(z_2-z_4)^{-1},$ and the four fermion correlator is
\eqn\fourspin{
\vev{s_{\th^j}(z_1)\ s_{-\th^j}(z_2)\ s_{\nu^j}(z_3)\ s_{-\nu^j}(z_4)}=
\fc{(z_1-z_2)^{-(1-\th^j)^2}}{(z_3-z_4)^{\, (1-\nu^j)^2}}\ 
\lf(\fc{z_{13}z_{24}}{z_{14}z_{23}}\ri)^{(1-\th^j)(1-\nu^j)}\hskip-.4cm.}

The bosonic twist correlators $Z_b=\vev{\sigma_{-\th^j}(z_1)\ \sigma_{\th^j}(z_2)\ \sigma_{-\nu^j}(z_3)\ 
\sigma_{\nu^j}(z_4)}$
are slightly more involved.
The four twist correlation function can be computed by adapting 
the arguments of \doubref\HV\DFMS\ to the open string case.
In particular the result for generic fluxes/angles, but $\nu=\theta$,
has been obtained in \GNS. Since all twist fields 
are inserted at the boundary of the disk, the correlator for
$\nu\neq \theta$ can be read off from the closed string correlator 
\doubref\BKM\EJSS\ by ``taking the square root'':\foot{See 
also \threeref\david\cvetic\abel\ for a further discussion of the open string
case.}  
\eqn\foursigma{
Z_b=\Cc_j \,(-1)^{2h_{\si_{\th^j}}+2h_{\si_{\nu^j}}}
z_{12}^{-2h_{\si_{\th^j}}}\ z_{34}^{-2h_{\si_{\nu^j}}}
\ I_j(x)^{-1/2}\ Z^{cl}_j\lf(\fc{z_{13}z_{24}}{z_{14}z_{23}}\ri)^{\h\th^j+\h\nu^j-\th^j\nu^j}\ ,} 
where  $\CC_j=\sqrt{\sin(\pi\theta)}$, 
$x=z_{12}z_{34}/z_{13}z_{24}$ and $Z^{cl}_j$ denotes the instanton
sum \GNS. Moreover
\eqn\II{
I_j(x)=\fc{1}{\pi}\sin(\pi\th^j)\ [B_{1;j}\ G_{2;j}(x)\ H_{1;j}(1-x)+B_{2;j}\ G_{1;j}(x)\ 
H_{2;j}(1-x)]\ ,}

\noindent
where $B_{1;j}=\fc{\Gamma(\th^j)\Gamma(1-\nu^j)}{\Gamma(1+\th^j-\nu^j)}$, 
$B_{2;j}=\fc{\Gamma(\nu^j)\Gamma(1-\th^j)}{\Gamma(1+\nu^j-\th^j)}$,
$G_{1;j}(x)=\ _2F_1[\th^j,1-\nu^j,1;x]$, $G_{2;j}(x)=\ _2F_1[1-\th^j,\nu^j,1;x]$,
$H_{1;j}(x)=\ _2F_1[\th^j,1-\nu^j,1+\th^j-\nu^j;x]$, and 
$H_{2;j}(x)=\ _2F_1[1-\th^j,\nu^j,1-\th^j+\nu^j;x]$.
In total the contribution of the internal twist fields amounts to
\eqn\amount{
(-1)^{-\sum\limits_{j=1}^3(\th^j)^2+(\nu^j)^2}\ \fc{z_{13}z_{24}}{z_{12}z_{14}z_{23}z_{34}}
\prod_{j=1}^3 \Cc_j\ \ I_j(x)^{-1/2}\, Z^{cl}_j}
due to \susy.
Putting everything together, the amplitude \todoi\ becomes
\eqn\total{\eqalign{
\Ac_{\A\oA\AA\oAA}(k_i)&=\fc{t}{2}\  
\int^\infty_{-\infty}\ \prod\limits_{i=1}^4 dz_i\ V_{\rm CKG}^{-1}\  
\prod_{j=1}^3 \Cc_j\ I_j(x)^{-\h}\, Z^{cl}_j\cr
&\times z_{12}^{-\al' s-1}\ z_{13}^{-\al' t}\ z_{14}^{-\al' u-1}\ z_{23}^{-\al' u-1}\ 
z_{24}^{-\al' t}\ z_{34}^{-\al' s-1}\ .}}
We fix three vertex positions according to \positionfour\ and introduce the factor
$z_\infty^2$ to account for the $c$--ghost correlator. 
With that choice \total\ reduces to:
\eqn\totali{
\Ac_{\A\oA\AA\oAA}(k_i)=\fc{t}{2}\ 
\int^\infty_{-\infty}\  dz\  z^{-\al' s-1}\ (z-1)^{-\al' u-1}
\prod_{j=1}^3  \Cc_j\ I_j(z)^{-\h}\, Z^{cl}_j\ .}

In order to extract the Yukawa couplings, we investigate the limit $z\rightarrow \infty $ 
in the integrand of the four matter fields amplitude \totali.
After using the relation \doubref\BKM\EJSS
\eqn\tutzing{
\lim_{z\ra\infty}\ \pi^{-1}\sin(p\pi\th^j)\ I_j(z)^{-1}\lra\cases{
(-1)^{k-l}\ z^{\th^j+\nu^j}\ \Gamma_{\th^j,\nu^j}  
&$0<\th^j+\nu^j<1$\ ,\cr
-(-1)^{k-l}\ z^{2-\th^j-\nu^j}\ \Gamma_{1-\th^j,1-\nu^j} 
& $1<\th^j+\nu^j<2$\ ,}}
with
\eqn\tum{
\Gamma_{\th^j,\nu^j}=\fc{\Gamma(1-\th^j)\ \Gamma(1-\nu^j)\ \Gamma(\th^j+\nu^j)}{\Gamma(\th^j)\ 
\Gamma(\nu^j)\ \Gamma(1-\th^j-\nu^j)}\ ,}
the integrand of \totali\ becomes in the limit $z\ra\infty$:
\eqn\becomes{\fc{t}{2}\ 
\prod_{j=1}^3  \Gamma_{\th^j,\nu^j}^{1/2}\ |W_j|^2\ \int^\infty dz\  z^{-\al' t-1}
\lra\ \prod_{j=1}^3  \Gamma_{\th^j,\nu^j}^{1/2}\ |W_j|^2.}
Here the $W_j$ is the superpotential describing the (classical) 
world--sheet disk instanton contributions elaborated in \doubref\CIM\cvetic. 
  From the above one finds the relative 
factor 
\eqn\Yukawa{
\prod_{j=1}^3     \lf[
\fc{\Gamma(1-\th^j)\ \Gamma(1-\nu^j)\ \Gamma(\th^j+\nu^j)}{\Gamma(\th^j)\ 
\Gamma(\nu^j)\ \Gamma(1-\th^j-\nu^j)}\ri]^{1/4}\ .}
between the Yukawa couplings in the string and the field theory basis, 
in agreement with the results of section 5.
Note that the correct power $1/4$ of the above factor 
is in contrast to a power $1/2$ claimed in Refs. 
\doubref\cvetic\AbelYX.\foot{As mentioned already, 
the quantum part encoded in $\Gamma_{\th^j,\nu^j}$
is at any rate just the square root of the closed 
string calculation, and can therefore be read off from 
\eqq (4.16) of \EJSS.}

\newsec{Conclusions}

For \tb orbifold/orientifold compactifications with $D9, D7, D5$ and $D3$--branes with internal
background fluxes we have calculated the complete 
(bosonic) tree--level action up to second order in the space--time momenta and scalar 
matter fields. Our findings directly apply to \ti orbifold/orientifold compactifications 
with $D9$ and $D5$--branes. Moreover, after performing $T$--duality \duality, our deliverables
take over to \ta orbifold/orientifold compactifications with intersecting $D6$--branes.
Although our  outcomes hold also for the non--supersymmetric case, here let us review those
for the N=1 supersymmetric case in $D=4$. 
We collect the main formulae for \tb orbifold/orientifold compactifications 
with $D9$--branes with non--trivial background fluxes.

In section 3 the holomorphic gauge kinetic function $f$ has been determined via disk scattering of two gauge fields and one modulus\foot{One should 
mention at this point that the string computation of \GNS\ already implicitly
contained this result.} with the result \susygaugenine:
\eqn\summerf{
f(s,t^j)=|n^1 n^2 n^3|\ \lf(s -\ap^{-2}f^1f^2\ t^3-\ap^{-2}f^1f^3\ t^2-\ap^{-2}f^2f^3\ t^1\ri)\ .}
Here, $f^j$ are three background fluxes on the $D9$--brane, referring to the three
tori $T^{2,j}$ the $D9$--brane is wrapped on with the wrapping numbers $n^j$. 

The Yukawa couplings and the superpotential were derived by computing open
string four point amplitudes on the disk.
Further details may be looked up in section 6.

The K\"ahler potential for the untwisted fields 
has been derived in section 4 (see also the appendix) and 
is summarized in \puttogether, while the metric for the general case 
has been determined in section 5. For completeness we discuss here also the
special case of matter fields with one angle equal to zero, 
which appear in the context the T-dual of pure $2p$-branes. 
Although the metric 
for these cases does not, in general, follow from the generic formula 
due to the appearance of extra massless states, the computation can
be straightforwardly adapted. For pure 5 and 9 branes one finds
\eqn\htmm{
G_{C^{95_i}\ov C^{95_i}=}\fc{1}{(t_1^j t_1^k u_1^ju_1^k)^\h},\qquad
G_{C^{5_i5_j}\ov C^{5_i5_j}}=\fc{1}{(s_1 t_1^k u_1^iu_1^j)^\h},}
which again agrees with the expressions of \AFIV\ in the case of fixed
complex structure.

We have determined the K\"ahler potential for the metric moduli $T^j,U^j$ and matter field moduli
$C_i,C_\th$ in the absence of Wilson line moduli.
It would be very interesting to extend our calculation to the case, when Wilson lines
are turned on.
The moduli space of metric and matter fields of $D=4$ Calabi--Yau compactifications
without branes shows some restricted structure, known as special geometry \DKL.
This is the case for heterotic N=1 Calabi--Yau compactifications and follows 
from the underlying $(2,2)$ superconformal world--sheet symmetry, which governs the closed string
vertex operators.
A natural question is, whether the geometry of the moduli space of 
N=1 \ti or type $II$ compactifications involving both open 
and closed strings (and $D$--branes) shows a similar restricted structure.
To address this questions, one
has to calculate four point--disk scattering amplitudes involving both open and closed strings 
to obtain conditions on the Riemann tensor of the moduli space.

Finally, the discrete symmetries acting on the moduli fields imply 
non--trivial transformations on the various vertex operators for the metric and matter field 
moduli, quite similar to the heterotic case \threeref\FLT\LMN\EJSS.
It would be interesting to study those transformation properties within these \ti and type $II$
vacua, used in this paper.

\bigskip
\centerline{\bf Acknowledgments }\nobreak
\bigskip
We are grateful to Niels Bernhardt for participation in an early stage of this project.
In addition, we thank Susanne Reffert for many useful remarks and 
Jan Louis for discussion.
This work is supported in part by the Deutsche 
Forschungsgemeinschaft (DFG), and the German--Israeli Foundation (GIF).

\appendix\appA{Scattering of two closed string moduli off the brane} 

Finally, in this section we shall determine the function $\hat{K}(M,\ov M)$ in \eqq \kaehler\ 
through string  amplitudes.
This term describes the metric of the K\"ahler and complex structure moduli $T^j,U^j$, 
respectively.
We consider the scattering of two closed string moduli fields $T^j,U^j$ on the disk.
We perform our calculation on the \tb side with general fluxes $\Fc^j$ on the $D9$--brane
turned on.

\subsec{Disk amplitudes with two closed strings}

Following the notation introduced in section 2 we consider the case $N_o=0,\ N_c=2$.
As we have already reviewed there,  a closed string vertex operator inserted on the disk 
${\cal H}_+$
is split into two open string vertex operators depending respectively holomorphically or
anti--holomorphically on the complex sphere coordinate $z$. Their interactions are described
by the correlator \greeni\ with the matrix $D^{ij}$.
Again the closed string momentum $q$ is distributed to the two open strings according to 
\closedm. Hence, two closed string vertex operators with momenta $q_1,q_2$ 
on the disk are treated as four
open strings of momenta $\h q_1,\h Dq_1,\h q_2$ and $\h Dq_2$, respectively.
Only the momentum parallel to the $D$--brane is conserved:
\eqn\mconservec{
q^{||}_1+q^{||}_2=0\ .}
After introducing the four momenta $k_i$ entering in the four open string vertex operators
\eqn\momentumcl{
k_1=q_1\ \ ,\ \ k_2=Dq_1\ \ ,\ \ k_3=q_2\ \ ,\ \ k_4=Dq_2}
\eqq \mconservec\ translates into \conserve, with $k_i^2=0$. 
Again, we may introduce the Mandelstam variables \mandelstamm, with $s+t+u=0$.  
However, now \eqq \mandelstamm\ implies the constraints \HK
\eqn\consc{
s=2\ (q_1^{||})^2=2\ (q_2^{||})^2=-2\ q^{||}_1\ q^{||}_2\ \ \ \ ,\ \ \ t=q_1q_2\ ,}
in contrast to \cons\ for the case $N_o=2,\ N_c=1$.

\subsec{Disk scattering of two metric moduli fields}

We shall first consider the scattering of two K\"ahler moduli $T^i$ and $\ov T^j$ on the disk:
\eqn\considerclosed{
\Ac_{T^i\ov T^j}(q_1,q_2)=C\int  d^2z_1\ d^2z_2\ V_{\rm CKG}^{-1}\ \vev{
V^{(0,0)}_{T^i}(\ov z_1,z_1,q_1)\ V^{(-1,-1)}_{\ov T^j}(\ov z_2,z_2,q_2)}\ .}
There are two contributions, one $X_1$ coming from the contraction of only two internal 
fermions and a second $X_2$ from contracting all four internal fermions:
\eqn\twocontr{\eqalign{
\Ac_{T^i\ov T^j}(q_1,q_2)&=C\ \fc{1}{(T^i-\ov T^i)\ (T^j-\ov T^j)}\ 
\int  d^2z_1\ d^2z_2\ V_{\rm CKG}^{-1}\ \fc{\Ec}{\ov z_2-z_2}\cr
&\hskip-0.9cm\times \lf\{\fc{D^j\ \ov{D^i}}{(\ov z_1-z_1)^2(\ov z_2-z_2)}
-\fc{s}{\ov z_1-z_1}\ \lf[\fc{D^j\ \ov{D^i}}{(\ov z_1-z_1)(\ov z_2-z_2)}-
\fc{\delta^{ij}}{(\ov z_1-\ov z_2)(z_1-z_2)}\ri]\ri\}.}}
On the disk the conformal Killing volume $V_{\rm CKG}^{-1}$ is canceled by fixing three
positions and introducing the respective ghost correlator.
A convenient choice of vertex operator positions is \HKi
\eqn\Choice{
z_1=iy\ \ ,\ \ \ov z_1=-iy\ \ ,\ \ z_2=i\ \ ,\ \ \ov z_2=-i\ ,}
which implies the ghost factor $1-y^2$.
With this choice the amplitude $\Ac_{T^i\ov T^j}$ 
becomes\foot{With this choice \Choice, the exponential $\Ec$ in \expo\ becomes 
$\Ec=\lf|\fc{4y}{(1+y)^2}\ri|^s\ \lf|\fc{1-y}{1+y}\ri|^{2t}$.}:
\eqn\fertig{\eqalign{
\Ac_{T^i\ov T^j}(q_1,q_2)&=\fc{C}{4}\ \fc{1}{(T^i-\ov T^i)\ (T^j-\ov T^j)}\cr 
&\times \int_0^1  dy\ \lf[\fc{4y}{(1+y)^2}\ri]^s\ \lf(\fc{1-y}{1+y}\ri)^{2t}\ 
\lf[D^j\ \ov{D^i}\ \fc{1-s}{4}\ \fc{1-y^2}{y^2}-\delta^{ij}\ s\ \fc{1+y}{y(1-y)}\ri]\ .}}
Now we perform  the same steps as described in \HKi\ to bring the above integral into
the standard form \veneziano:
\eqn\TTc{\eqalign{
\Ac_{T^i\ov T^j}(q_1,q_2)&=\fc{1}{(T^i-\ov T^i)\ (T^j-\ov T^j)}\ \lf[D^j\ \ov{D^i}\ (s-1)\ V(t,s-2)
+\delta^{ij}\ s\ V(t-1,s-1)\ri]\cr
&=\fc{1}{(T^i-\ov T^i)\ (T^j-\ov T^j)}\ \lf(D^j\ \ov{D^i}\ t+\delta^{ij}\ s\ri)\ (1-s)\ 
\fc{\Gamma(t)\Gamma(s-1)}{\Gamma(s+t)}\ .}}
With the identity $D^i\ov{D^i}=1$ we obtain for the case $i=j$
\eqn\mi{\eqalign{
\Ac_{T^i\ov T^i}(q_1,q_2)&=
-\fc{1}{(T^i-\ov T^i)^2}\ (s+t)^2\ \fc{\Gamma(s)\ \Gamma(t)}{\Gamma(1+s+t)}\cr
&=-\fc{1}{(T^i-\ov T^i)^2}\ \lf(1-\fc{\pi^2}{6}st\ri)\ \fc{(s+t)^2}{st}+\Oc(k^6)\ ,}}
whereas the case $i\neq j$ gives:
\eqn\mii{\eqalign{
\Ac_{T^i\ov T^j}(q_1,q_2)&=-\fc{D^j\ \ov{D^i}}{(T^i-\ov T^i)(T^j-\ov T^j)}\   
\fc{\Gamma(s)\ \Gamma(t+1)}{\Gamma(s+t)}\cr
&=-\fc{D^j\ \ov{D^i}}{(T^i-\ov T^i)(T^j-\ov T^j)}\  \lf(1-\fc{\pi^2}{6}st\ri)\ \fc{s+t}{s}+
\Oc(k^6)\ .}}
Note, that the first result \mi\ is completely independent on the value of the fluxes $f^j$,
since the parameter $D^i$ has dropped out. 
Only in that case we find a term proportional to $s/t$ to conclude
\eqn\metricconc{
G_{T^i\ov T^j}=-\fc{\delta^{ij}}{(T^i-\ov T^i)^2}\ .}
That pole--term describes a pole on the sphere with the closed string vertices coming close
together. 
Since the parameters $D_i, \ov{D_i}$, which determine the mixing of the open string 
boundary conditions on the $D9$--brane, have completely dropped out for $\Ac_{T^i\ov T^i}$, 
the result is independent, whether the internal open string fields of the vertex operators 
$V_{T^i}$ and $V_{\ov T^i}$ obey Dirichlet, Neumann or mixed boundary conditions.
This means, that we would have anticipated the same result for $\Ac_{T^i\ov T^i}$
from scattering $T^i$ and $\ov T^i$ in the presence of a $D5$--brane wrapped around a torus 
$T^{2,k}$, with $k\neq i$. In that case the torus $T^{2,i}$ would be transversal to
the $D5$--brane with the open string fields obeying Dirichlet boundary conditions.

Let us now repeat our previous calculation for the disk scattering of two complex structure moduli
$U^i$ and $\ov U^j$. The details are similar to the above derivation. Again, one has to 
use the identity $\ov{D^i}D^i=1$ to arrive at:
\eqn\UUc{
\Ac_{U^i\ov U^j}(q_1,q_2)=-\fc{\delta^{ij}}{(U^i-\ov U^i)^2}\ s\ 
\fc{\Gamma(t)\Gamma(s+1)}{\Gamma(1+s+t)}=\fc{\delta^{ij}}{(U^i-\ov U^i)^2}\ 
\fc{s}{t}\ \lf(1-\fc{\pi^2}{6}st\ri)+\Oc(k^6)\ .}
 From this amplitude one concludes, that there is no coupling between complex structure moduli
from different planes. Furthermore the K\'ahler metric $G_{U^i\ov U^i}$ for the complex 
structure moduli is encoded in the pole:
\eqn\UUC{
G_{U^i\ov U^j}=-\fc{\delta^{ij}}{(U^i-\ov U^i)^2}\ .}

Finally, the disk scattering of one K\"ahler modulus $T^i$ and one complex structure modulus
$\ov U^j$ gives a vanishing result due to internal charge conservation:
\eqn\TUc{
\Ac_{T^i\ov U^j}(q_1,q_2)=0\ .}

The above scattering results, \ie the metrics  \metricconc, \UUC\ and \TUc, 
can be derived formally\foot{It should be kept in mind, that the scalars 
$S,T^j$ and $U^j$ do not represent scalars of N=1 chiral multiplets.} 
from K\"ahler potential for the metric moduli:
\eqn\kaehlerconclude{
-\ln(S-\ov S)-\sum_{j=1}^3 \ln (T^j-\ov T^j)-\sum_{j=1}^3 \ln (U^j-\ov U^j)\ .}
The first term describes the kinetic energy term for the dilaton field $S$, with
$\im(S)=e^{-2\phi_4}$ (\cf Appendix \appB).
Alternatively, in terms of the {\it physical} moduli fields, introduced in
\eqqs \gaugeDnine, \gaugeDfive\ and \fieldtypea, the K\"ahler potential $\hat K$ 
without the matter field metrics may be also written (\cf \kpstu):
\eqn\Kaehlerconclude{
\hat{K}(M,\ov M)=-\kappa_4^{-2}\ \ln(s+\ov s)-\kappa_4^{-2}\ \sum_{j=1}^3 \ln (t^j+\ov t^j)-
\kappa_4^{-2}\ \sum_{j=1}^3 \ln (u^j+\ov u^j)\ .}
To conclude, the K\"ahler potential \Kaehlerconclude\ takes the same form as 
in ordinary \ti or \tb orientifold/orbifold compactifications without fluxes.
For those models without fluxes \eqq \Kaehlerconclude\ has been derived based on symmetry 
arguments in \AFIV. The fields entering in 
\Kaehlerconclude\ stem from the untwisted sector, which explains its quite universal 
appearance even in the presence of fluxes. 
As we have seen in our calculations, it essentially makes no difference, whether we consider
disk scattering of K\"ahler and complex structure moduli fields in the presence of a $D9$--brane 
or $D5$--brane. In other words, the result is independent, whether the torus, to which the
moduli fields refer, is parallel or transverse to the $Dp$--brane.
In fact, we could just stick to the bulk of the $Dp$--brane, where only the closed string sector 
of the underlying \tb orbifold/orientifold lives. From scattering of four moduli fields on the 
sphere one ends up with the same result \Kaehlerconclude.
It has been already pointed out in \DKL,
that $D=4$ heterotic N=1 or N=2 type $II$ compactification on the same manifold have the same
metric moduli spaces up to second order in the momenta. We have checked this
for the kind of models we discuss in this article.

\appendix\appB{Matter and gauge fields coupling to the graviton and dilaton}

In this section we shall be interested in the dilaton dependence of the 
gauge couplings and matter field metrics, which have been the subject of
the previous sections.
This discussion is necessary in order to obtain the dilaton dependence of the metrics.
The vertex operator for the bosonic massless closed string modes describing a graviton, 
dilaton or anti--symmetric tensor in the $(-1,-1)$ ghost picture is given 
by:\foot{For further details, see also \MSD.}
\eqn\dilaton{
V_G^{(-1,-1)}(\ov z,z,q)=\epsilon_{\mu\nu}\ e^{-\tilde\phi(\ov z)}\ 
e^{-\phi(z)}\ \tilde\psi^\mu(\ov z)\ \psi^\nu(z)\ e^{iq_\nu X^\nu(\ov z,z)}\ .}
The polarization tensor $\eps_{\mu\nu}$ is subject to the on--shell conditions 
$\eps_{\mu\nu} q^\mu=0=\eps_{\mu\nu} q^\nu$ and $q^2=0$.
Apart from these constraints we shall perform the calculation for arbitrary polarization 
$\epsilon_{\mu\nu}$, thus allowing to also extract the gauge and matter field couplings to 
the graviton and anti--symmetric tensor.
The polarization $\epsilon_{\mu\nu}$ in \dilaton\ determines the relevant closed string 
state\foot{Note, that in \ti or \tb orientifolds the space--time anti--symmetric
tensor $b_{\mu\nu}$ is generically projected out.}:
\eqn\polarization{\eqalign{
\epsilon_{\mu\nu}=\epsilon_{\nu\mu}\ \ \ ,\ \ \ & {\rm Graviton\ ,}\cr
\epsilon_{\mu\nu}=-\epsilon_{\nu\mu}\ \ \ ,\ \ \ &{\rm Kalb-Ramond\ ,}\cr
\epsilon_{\mu\nu}=\fc{1}{\sqrt 2}\ (\eta_{\mu\nu}-q_\mu \ov q_\nu-q_\nu \ov q_\mu)\ \ \ ,\ \ \ &
\ov q^2=0\ \ ,\ \ q_\mu\ov q^\mu=1\ ,\ \ \ {\rm Dilaton\ .}}}
We consider disk amplitudes with one massless bosonic supergravity $NS$ field inserted in the bulk 
and two open string states, inserted at the boundary of the disk. The open string states
are respectively untwisted matter fields $C_i$, twisted matter fields $C_\th$, and 
gauge fields $A^a_\mu$.
First, we shall calculate the string $S$--matrix:
\eqn\first{
\Ac_{C_i\ov C_i G}=\fc{-i}{\pi}\ \int \fc{dz_1 dz_2 d^2z_3}{V_{\rm CKG}}\ 
\vev{V_{C_i}^{(0)}(z_1,k_1)\ V_{\ov C_i}^{(0)}(z_2,k_2)\ V_{G}^{(-1,-1)}(\ov z_3,z_3,q)}\ .}
The untwisted matter field vertex operators have been given in \zeromatter.

In \first\ two contractions are possible: The first, denoted by $X_1$, with the two internal bosons
from the matter vertices contracted and a second $X_2$ with their two internal fermions
contracted. All correlators are basic and may be looked up from section 2.
To this end we find
\eqn\firstd{
\Ac_{C_i\ov C_i G}=\fc{-i}{\pi}\ \int \fc{dz_1 dz_2 d^2z_3}{V_{\rm CKG}}\ \Ec\ (X_1+X_2)\ ,}
with
\eqn\firstdd{\eqalign{
X_1&=-\Tr(\epsilon)\ \fc{1}{(z_1-z_2)^2}\ \fc{1}{(\ov z_3-z_3)^2}\ ,\cr
X_2&=\fc{1}{(z_1-z_2)(\ov z_3-z_3)}\ 
\lf[\fc{s\ \Tr(\epsilon)}{(z_1-z_2)(\ov z_3-z_3)}-
\fc{k_1\eps k_2}{(z_1-\ov z_3)(z_2-z_3)}+\fc{k_1\eps^t k_2}{(z_1-z_3)(z_2-\ov z_3)}\ri],}}
and the correlator \expo.
After fixing the vertex positions like \position\ and introducing the ghost correlator
\cghost, we obtain:
\eqn\firstddd{\eqalign{
\Ac_{C_i\ov C_i G}&=2^{-4t-1}\ \pi^{-1}\lf[\ \fc{1}{4}\ \Tr(\epsilon)\ (1+2t)\ I(-1-2t,0)
+i\ k_1^t(\epsilon+\epsilon^t) k_2\ I(-2t,1)\ \ri]\cr
&=2^{-2t-1}\ \pi^{-1/2}\lf[\ \fc{1}{2}\ \Tr(\epsilon)\ (1+2t)\ \fc{\Gamma(-\h-t)}{\Gamma(-t)}
-\ k_1^t(\epsilon+\epsilon^t) k_2\ \fc{\Gamma(\h-t)}{\Gamma(1-t)}\ \ri]\cr
&=\h\ \lf[t\ \Tr(\epsilon)-t\ k_1^t(\epsilon+\epsilon^t) k_2\ +\Oc(t^2)\ \ri]\ .}}
The first term gives rise to the coupling of the two untwisted matter fields $C_i,\ov C_i$
to the dilaton, while the second term describes their coupling to the graviton (at zero momentum).
Obviously, there is no coupling of the anti--symmetric tensor $b_{\mu\nu}$ to the matter fields.

Next, we calculate the disk amplitude
\eqn\First{
\Ac_{C_\th\ov C_\th G}=\fc{-i}{\pi}\ \int \fc{dz_1 dz_2 d^2z_3}{V_{\rm CKG}}\ 
\vev{V_{C_\th}^{(0)}(z_1,k_1)\ V_{\ov C_\th}^{(0)}(z_2,k_2)\ V_{G}^{(-1,-1)}(\ov z_3,z_3,q)}\ .}
involving two twisted matter fields $C_\th,\ov C_\th$ inserted at the boundary of the disk
and one closed string modulus from the bulk. The latter describes a massless bosonic 
member of the supergravity multiplet. It proves to be convenient to work with the matter vertices
in the zero--ghost picture \mattervertexi\ and the closed string vertex \dilaton.
Since, the bosonic and fermionic twist fields in \First\ completely decouple from the 
closed string vertex they correlators may be determined independently.
Due to internal charge conservation they boil down to products of the basic Green's functions
\basic\ and \basici:
\eqn\boil{\eqalign{
&\vev{[\sum_{l=1}^3 \tau_{-\th^l}\ t_{\th^l}(z_1)\prod^3_{j=1\atop j\neq l} 
\sigma_{-\th^j}\ s_{\theta^j}(z_1)]\ 
[\sum_{m=1}^3 \tau_{\th^{m}}\ t_{-\th^{m}}(z_2)
\prod^3_{k=1\atop k\neq m}\sigma_{\th^{k}}\ 
s_{-\theta^{k}}(z_2)]}=\fc{1}{(z_1-z_2)^2}\ ,\cr
&\vev{[\prod^3_{j=1} \sigma_{-\th^j}\ s_{\theta^j}(z_1)]\ 
[\prod^3_{k=1}\sigma_{\th^{k}}\ s_{-\theta^{k}}(z_2)]}=\fc{1}{z_1-z_2}}}
The contraction of the space--time fields in \First\ are the same as in \first. In fact,
it is not hard to see, that the amplitude 
$\Ac_{C_\th\ov C_\th G}$ takes the same form as \firstd\ and \firstdd. Hence we conclude:
\eqn\Firstddd{
\Ac_{C_\th\ov C_\th G}=\Ac_{C_i\ov C_i G}\ .}

Finally, the coupling of the dilaton to two gauge fields at the boundary of the disk
may be directly taken from \HK:
\eqn\FirstdddD{\eqalign{
\Ac_{A^aA^a G}&=-2\ \Tr(\eps)\ \lf[\ (p_1\xi_2)(p_2\xi_1)-(p_1p_2)(\xi_1\xi_2)\ \ri]\ t\ 
\fc{\Gamma(-2t)}{\Gamma(1-t)^2}\cr
&=\Tr(\eps)\ \lf[\ (p_1\xi_2)(p_2\xi_1)-(p_1p_2)(\xi_1\xi_2)\ \ri]\ \lf[1+\Oc(t)\ri]\ .}}

Hence, we conclude, that both the matter metrics and the gauge couplings have
universal couplings to the dilaton at disk tree--level.

\appendix{\appC}{Disk scattering with excited twist fields}

Let us now calculate the string $S$--matrix
\eqn\todoapp{
\Ac_{\A\oA U^j}=\int \fc{dz_1 dz_2 d^2z_3}{V_{\rm CKG}}\ 
\vev{V_\A^{(0)}(z_1,k_1)\ V_{\oA}^{(0)}(z_2,k_2)\ V_{U^j}^{(-1,-1)}(\ov z_3,z_3,q)}}
on the disk to extract information on the metric of two matter fields.
Furthermore, the two matter field vertices are inserted at the boundary
of the disk, \ie $z_1=\ov z_1,z_2=\ov z_2$, while the closed string 
vertex operator for the $U^j$--modulus, given in \vertexTU, is inserted in the bulk. 
The latter is chosen in the $(-1,-1)$ ghost picture, while the matter vertices are taken 
in the $0$ ghost picture \mattervertexi\ in order to guarantee a total ghost 
charge of $-2$ on the disk.
Due to internal charge conservation, there are two non--vanishing contributions
to the amplitude \todo: One from the contraction of the internal fermions of the $U^j$--modulus
with the both excited twist fields 

They arise from the contraction of the $U^j$--vertex operator
with either only the first terms of the matter vertices \mattervertexi\ 
or with only the second terms. We shall denote these two possibilities by $X_1$ and $X_2$, 
respectively.
The contraction of the exponentials is given in \expo, i.e by $\Ec$.
The correlators \basic, \basic\ are needed, together with the correlator:
\eqn\Xii{
\vev{t_{\th^j}(z_1)\ t_{-\th^j}(z_2)\ \tilde\Psi^j(\ov z_3)\ \ov\Psi^j(z_3)}=
\fc{\ov D^j}{\ov z_3-z_3}\ (z_1-z_2)^{-\th_j^2}\ 
\lf[\fc{(z_1-z_3)\ (z_2-\ov z_3)}{(z_1-\ov z_3)\ (z_2-z_3)}\ri]^{\th_j}}
for the first contraction, while for the second contraction we use \needcorri.
We obtain
\eqn\todoappp{
\Ac_{\A\oA U^j}=\fc{\ov D^j}{{U}^j-\ov {U}^j}\ 
\int \fc{dz_1 dz_2 d^2z_3}{V_{\rm CKG}}\ \Ec\ (z_1-z_2)^{-2}
(\ov z_3-z_3)^{-2}\ \lf(X_1-2t\ X_2\ri)\ ,}
with:
\eqn\obtaini{\eqalign{
X_1&=(1-\th^j)\ \lf[\fc{(\ov z_3-z_1)\ (z_3-z_2)}{(z_3-z_1)\ (\ov z_3-z_2)}\ri]^{-\th^j}+
\th^j\ \lf[\fc{(\ov z_3-z_1)\ (z_3-z_2)}{(z_3-z_1)\ (\ov z_3-z_2)}\ri]^{1-\th^j}\ ,\cr
X_2&=-\ \lf[\fc{(\ov z_3-z_1)\ (z_3-z_2)}{(z_3-z_1)\ (\ov z_3-z_2)}\ri]^{1-\th^j}\ .}}
The two terms $X_1$ and $X_2$ agree with the expressions, given in \zusatz.
Hence, the string $S$--matrix \todoapp\ gives the same result as the amplitude \todo.

\listrefs
\end